%% file: synchronization_fine_freq_time_varying_main_v6.tex
\documentclass[twocolumn,journal]{IEEEtran}
\usepackage[T1]{fontenc}
\usepackage[latin9]{inputenc}
\usepackage{array}
\usepackage{verbatim}
\usepackage{prettyref}
\usepackage{float}
\usepackage{amsmath}
\usepackage{amsthm}
\usepackage{amssymb}
\usepackage{graphicx}
\usepackage[unicode=true,
 bookmarks=true,bookmarksnumbered=true,bookmarksopen=true,bookmarksopenlevel=1,
 breaklinks=false,pdfborder={0 0 0},pdfborderstyle={},backref=false,colorlinks=false]
 {hyperref}
\hypersetup{pdftitle={Your Title},
 pdfauthor={Your Name},
 pdfpagelayout=OneColumn, pdfnewwindow=true, pdfstartview=XYZ, plainpages=false}

\makeatletter

\providecommand{\tabularnewline}{\\}
\floatstyle{ruled}
\newfloat{algorithm}{tbp}{loa}
\providecommand{\algorithmname}{Algorithm}
\floatname{algorithm}{\protect\algorithmname}

\theoremstyle{plain}
\newtheorem{thm}{\protect\theoremname}
\theoremstyle{definition}
\newtheorem{problem}[thm]{\protect\problemname}
\theoremstyle{remark}
\newtheorem{rem}[thm]{\protect\remarkname}

\usepackage[caption=false,font=footnotesize]{subfig}
\usepackage{cite}

\makeatother

\providecommand{\problemname}{Problem}
\providecommand{\remarkname}{Remark}
\providecommand{\theoremname}{Theorem}

\begin{document}
\title{Maximum a Posteriori Probability (MAP) Joint Carrier Frequency Offset
(CFO) and Channel Estimation for MIMO Channels with Spatial and Temporal
Correlations}
\author{Ibrahim Khalife, Ali Abbasi, Zhe Feng, Mingda Zhou, Xinming Huang,
Youjian Liu \thanks{The work is partially supported by NSF-2128659, 2203060. Zhou and
Huang are with the Department of Electrical and Computer Engineering,
Worcester Polytechnic Institute. Other authors are with the Department
of Electrical, Computer, \& Energy Engineering, University of Colorado
at Boulder, e-mail: \protect\href{mailto:eugeneliuieee@ieee.org}{eugeneliuieee@ieee.org}. }}
\maketitle
\begin{abstract}
\input{synchronization_fine_freq_time_varying_abstract_v4.tex}
\end{abstract}


\input{synchronization_fine_freq_time_varying_Intro_v4.tex}

\input{synchronization_fine_freq_time_varying_notation_v4.tex}

\input{synchronization_fine_freq_time_varying_body_v6.tex}

\input{synchronization_fine_freq_time_varying_Results_v4.tex}

\input{synchronization_fine_freq_time_varying_conclusion_v4.tex}

\appendices{}

\input{synchronization_fine_freq_time_varying_apdx_v6.tex}

%\bibliographystyle{IEEEtran}
%\bibliography{Fine_Frequency_Synchronization,liu_research}

\input{synchronization_fine_freq_time_varying_main_v6.bbl}

\end{document}

%% file: synchronization_fine_freq_time_varying_abstract_v4.tex
We consider time varying MIMO fading channels with known spatial and
temporal correlation and solve the problem of joint carrier frequency
offset (CFO) and channel estimation with prior distributions. The
maximum a posteriori probability (MAP) joint estimation is proved
to be equivalent to a separate MAP estimation of the CFO followed
by minimum mean square error (MMSE) estimation of the channel while
treating the estimated CFO as true. The MAP solution is useful to
take advantage of the estimates from the previous data packet. A low
complexity universal CFO estimation algorithm is extended from the
time invariant case to the time varying case. Unlike past algorithms,
the universal algorithm does not need phase unwrapping to take advantage
of the full range of symbol correlation and achieves the derived Bayesian
Cramér-Rao lower bound (BCRLB) in almost all SNR range. We provide
insight on the the relation among the temporal correlation coefficient
of the fading, the CFO estimation performance, and the pilot signal
structure. An unexpected observation is that the BCRLB is not a monotone
function of the temporal correlation and is strongly influenced by
the pilot signal structures. A simple rearrangement of the 0's and
1's in the pilot signal matrix will render the BCRLB from being non-monotone
to being monotone in certain temporal correlation ranges. Since the
BCRLB is shown to be achieved by the proposed algorithm, it provides
a guideline for pilot signal design.

%% file: synchronization_fine_freq_time_varying_Intro_v4.tex
\section{Introduction\label{sec:tv_Introduction}}

\begin{comment}
In the body, talk about sine(theta). Add details in the simulation
results discussion.
\end{comment}

\begin{comment}
ours can be used in both bursty mode and tracking mode
\end{comment}

Carrier frequency offset (CFO) estimator is a critical component of
a communication system. It also has applications in radar and sensing
systems. We aim for joint carrier frequency offset (CFO) and fading
channel coefficients estimation for multiple-input-multiple-output
(MIMO) flat fading channels that takes advantage of the spatial and
temporal correlations whenever they are available. The results can
be applied to orthogonal frequency division multiplexing (OFDM) systems
as well. It completes the work started in \cite{Liu_2019GCCG_MAPJointFrequencyChannelEstimationMIMOSystemsSpatialCorrelation,Liu_2021TSP_MaximumPosterioriProbabilityMAPJointFineFrequencyOffsetChannelEstimationMIMOSystemsChannelsArbitraryCorrelation},
which was limited to the case of time invariant channels%
\begin{comment}
The extension to the time varying case in this work fulfills the requests
by the reviewers of \cite{Liu_2021TSP_MaximumPosterioriProbabilityMAPJointFineFrequencyOffsetChannelEstimationMIMOSystemsChannelsArbitraryCorrelation}.
\end{comment}
. In \cite{Liu_2021TSP_MaximumPosterioriProbabilityMAPJointFineFrequencyOffsetChannelEstimationMIMOSystemsChannelsArbitraryCorrelation},
we proved that the joint maximum a posteriori (MAP) estimation of
the CFO and channel can be achieved separately, i.e., by a MAP estimation
of the CFO using the spatial channel statistics, followed by a minimum
mean square error (MMSE) estimation of the channel assuming the estimated
CFO is true. We designed a state-of-the-art universal CFO estimation
algorithm that takes advantage of the spatial statistics of the channel,
has low complexity, and has no error floor. We also derived Cramér-Rao
lower bound (CRLB) and Bayesian Cramér-Rao lower bound (BCRLB) and
showed that the impact of pilot signal structures on the performance.
It is found that the same methodology works and the theory, algorithms,
and bounds in \cite{Liu_2021TSP_MaximumPosterioriProbabilityMAPJointFineFrequencyOffsetChannelEstimationMIMOSystemsChannelsArbitraryCorrelation}
are extended to the case of time varying channels and some unexpected
observations of the performance limit as a function of the temporal
correlation are revealed.

The CFO estimation problem is a well investigated topic. In \cite{Fitz_1997IToC_FrequencyOffsetCompensationPilotSymbolAssistedModulationFrequencyFlatFading,Vitetta_1998ICL_FurtherResultsCarrierFrequencyEstimationTransmissionsFlatFadingChannels,Stoica_2001C_FrequencyOffsetEstimationFlatfadingChannels},
the maximum-likelihood (ML) estimator and its variants are given for
pilot aided communications in a time varying flat fading channels.
The approximation $\sin(z)\doteq z$ is used widely in literature
to approximately solve for a stationary point of the ML metric. It
only utilizes a small lag of the symbol correlation if one wants to
avoid phase unwrapping or time variation. Phase unwrapping is as hard
as CFO estimation itself. The approximation may not be asymptotically
accurate and causes error floor in MIMO channels with nonzero means
because $z$ may not tends to zero even with infinite SNR. The CRLB
of joint channel and CFO estimation for time invariant MIMO channel
is derived in \cite{Stoica_2003ITSP_ParameterEstimationMIMOFlatfadingChannelsFrequencyOffsets}.
In \cite{Tarokh_2008ITSP_BoundsAlgorithmsFrequencySynchronizationCollaborativeCommunicationSystems},
BCRLB is derived for single antenna relay networks where and the frequency
is assumed to be Gaussian distributed. %
{} To avoid repetition and due to space limit, the readers are referred
to \cite{Liu_2021TSP_MaximumPosterioriProbabilityMAPJointFineFrequencyOffsetChannelEstimationMIMOSystemsChannelsArbitraryCorrelation}
for detailed discussion of more literature, e.g., \cite{Stuber_2001IGTC2G'_SynchronizationMIMOOFDMSystems,Ng_2004ITSP_SemiblindChannelEstimationMethodMultiuserMultiantennaOFDMSystems,Wang_2005ITC_EMbasedIterativeReceiverDesignCarrierfrequencyOffsetEstimationMIMOOFDMSystems,Kuo_2006IToC_MaximumlikelihoodSynchronizationChannelEstimationOFDMAUplinkTransmissions,Li_2006ITC_OptimalTrainingSignalsMIMOOFDMChannelEstimationPresenceFrequencyOffsetPhaseNoise,Ng_2007ITC_JointSemiblindFrequencyOffsetChannelEstimationMultiuserMIMOOFDMUplink,Ko_2009IToVT_JointChannelEstimationSynchronizationMIMOOFDMPresenceCarrierSamplingFrequencyOffsets,Berbineau_2011IToVT_JointCarrierFrequencyOffsetFastTimeVaryingChannelEstimationMIMOOFDMSystems,Hari_2012CN2NCO_JointEstimationSynchronizationImpairmentsMIMOOFDMSystem,Moretti_2013IWCL_JointMaximumLikelihoodEstimationCFONoisePowerSNROFDMSystems,Orozco-Lugo_2013IWCL_CarrierFrequencyOffsetEstimationOFDMAUsingDigitalFiltering,Gao_2016ITWC_ComputationallyEfficientBlindEstimationCarrierFrequencyOffsetMIMOOFDMSystems}

The contributions of the work are as follows.
\begin{itemize}
\item We start with the general case of allowing different pairs of transmit
antenna and receive antenna to have different CFOs, which may happen
in distributed MIMO systems. We show that when the CFOs of the same
receive antenna are the same ($\boldsymbol{f}_{r,t}=\boldsymbol{f}_{r}$),
the joint MAP CFO and channel estimation is equivalent to a separate
MAP estimation of the CFO with only the channel statistical information,
followed by an MMSE estimation of the channel with the estimated frequency
offset substituted in. The MAP solution includes the ML solution as
a special case when we let the variances of the CFO or the channel
approach infinity.
\item For the case of $\boldsymbol{f}_{r,t}=\boldsymbol{f}_{r}$, we derive
the stationary point condition of the MAP CFO estimation and extend
the universal CFO estimation algorithm developed for the time invariant
channel cases \cite{Liu_2021TSP_MaximumPosterioriProbabilityMAPJointFineFrequencyOffsetChannelEstimationMIMOSystemsChannelsArbitraryCorrelation}
to time varying channel cases, producing the state-of-the-art algorithm.
For the case of the same CFO for all antennas ($\boldsymbol{f}_{r}=\boldsymbol{f}$),
the extended universal algorithm only needs searching small grids
and solving a linear equation iteratively to find the best stationary
point almost exactly, as demonstrated by numerical results that the
universal algorithm achieves the BCRLB and CRLB for a wide range of
SNR without error floor, from time correlated fading to independent
fading. Unlike past algorithms, no phase unwrapping is needed to utilize
the full range of symbol correlations for the estimation.
\item The Cramér-Rao Lower bound (CRLB) and Bayesian Cramér-Rao Lower bound
(BCRLB) are derived in closed form for the case of $\boldsymbol{f}_{r}=\boldsymbol{f}$,
as a function of the spatial and temporal correlation. The CRLB/BCRLB
reveal interesting relations between the space and time correlation
of the fading and the CFO estimation performance. Specifically, a
tiny decrease of temporal correlation coefficient causes significant
performance deterioration but the time diversity may benefit the performance
if the fading has nonzero means. Thus, the CRLB/BCRLB may be a \emph{non-monotone}
function of the time correlation coefficient, depending on the pilot
signal structure and SNR, a phenomenon not reported in prior literature.
The best performing pilot signal structure is a function of the temporal
correlation.
\end{itemize}
The rest of this paper is organized as follows. Section \ref{sec:tv_System-Model}
provides the system model. In Section \ref{sec:tv_The-Optimization-Problem},
the joint MAP estimation of CFO and channel is shown to be separable.
In Section \ref{sec:tv_Fine-Frequency-Synchronization}, the universal
frequency synchronization algorithms are presented. To analyze the
performance limit, CRLB/BCRLBs as design guidelines are derived in
Section \ref{sec:tv_Performance-Analysis}. In Section \ref{sec:tv_Simulation-Results},
we show how the temporal correlation and the pilot signal structure
affect the BCRLB. Simulation results of the proposed algorithm are
compared with the BCRLB and CRLB. Section \ref{sec:tv_Conclusion}
concludes.

%% file: synchronization_fine_freq_time_varying_notation_v4.tex
\emph{Notation Convention:} We use the notation convention in Table
\ref{tab:Notations}. The power of the notation lies in the ability
to specify the entries of a matrix as a function of the row and column
indexes. It is convenient for organizing variables with multiple indices
into a matrix or a vector or vectorizing a matrix. 

\begin{table}[h]
\caption{\label{tab:Notations}Notation Convention}

\centering{}%
\begin{tabular*}{0.95\columnwidth}{@{\extracolsep{\fill}}|>{\raggedright}m{0.2\columnwidth}|>{\raggedright}p{0.65\columnwidth}|}
\hline 
Notation & Meaning\tabularnewline
\hline 
\hline 
$x$, $\vec{x}$, $X$ & scalar, \emph{column} vector, matrix\tabularnewline
\hline 
$\boldsymbol{x}$, $\vec{\boldsymbol{x}}$, $\boldsymbol{X}$ & random variable, column random vector, random matrix\tabularnewline
\hline 
$[a_{x_{1},x_{2}}]_{x_{1},x_{2}}$ & a matrix whose element at $x_{1}$-th row and $x_{2}$-th column is
$a_{x_{1},x_{2}}$, e.g., $\left[\begin{array}{cc}
a_{1,1} & a_{1,2}\\
a_{2,1} & a_{2,2}
\end{array}\right]=\left[a_{i,j}\right]_{i,j}$ ; $\left[e^{-j2\pi\frac{i\cdot j}{n}}\right]_{i=0:n-1,j=0:n-1}$
is a \emph{DFT matrix}; $x_{1}$ or $x_{2}$ can be continuous variables\tabularnewline
\hline 
$[a_{x}]_{x,x}$ & a diagonal matrix whose element at $x$-th row and $x$-th column
is $a_{x}$, e.g., $\left[\begin{array}{cc}
a_{1} & 0\\
0 & a_{2}
\end{array}\right]=\left[a_{i}\right]_{i,i}$\tabularnewline
\hline 
$\left[A_{x_{1},x_{2}}\right]_{x_{1},x_{2}}$  & a block matrix whose block at $x_{1}$-th row and $x_{2}$-th column
is $A_{x_{1},x_{2}}$\tabularnewline
\hline 
$\left[\vec{a}_{x}\right]_{1,x}$ & a matrix whose $x$-th column is $\vec{a}_{x}$, e.g., $\left[\begin{array}{cc}
\vec{a}_{1} & \vec{a}_{2}\end{array}\right]=\left[\vec{a}_{i}\right]_{1,i}$\tabularnewline
\hline 
$[a_{x}]_{x}$ & a column vector whose element at the $x$-th row is $a_{x}$, e.g.,
$\left[\begin{array}{c}
a_{1}\\
a_{2}\\
a_{3}
\end{array}\right]=\left[a_{i}\right]_{i}$\tabularnewline
\hline 
$\left[\vec{a}_{x}\right]_{x}$ & a tall vector whose $x$-th row of vector is $\vec{a}_{x}$, e.g.,
$\left[\begin{array}{c}
a_{1,1}\\
a_{2,1}\\
a_{1,2}\\
a_{2,2}
\end{array}\right]=$ $\left[\begin{array}{c}
\left[a_{i,1}\right]_{i}\\
\left[a_{i,2}\right]_{i}
\end{array}\right]=\left[\left[a_{i,j}\right]_{i}\right]_{j}$\tabularnewline
\hline 
$\left[\vec{a}_{x}^{T}\right]_{x}$  & a matrix whose $x$-th row is $\vec{a}_{x}^{T}$, e.g., $\left[\begin{array}{c}
\vec{a}_{1}^{T}\\
\vec{a}_{2}^{T}
\end{array}\right]=\left[\vec{a}_{i}^{T}\right]_{i}$\tabularnewline
\hline 
\end{tabular*}
\end{table}

%% file: synchronization_fine_freq_time_varying_body_v6.tex
\section{System Model\label{sec:tv_System-Model}}

\begin{comment}
\$result can be used in ofdm as it is flat
\end{comment}

We investigate joint CFO and time varying flat fading channel estimation
for MIMO systems. The result can be directly applied to OFDM system
as the channel for each sub-carrier is flat. We start with general
model that allows the CFO to be a function of both the transmit and
receive antenna indexes, later restricting it to be a function of
only the receive antenna index, and a function of neither, in order
to demonstrate where and what restrictions are needed for certain
results to hold true and for certain algorithms to work. The transmitter
has $l_{\text{t}}$ antennas and the receiver has $l_{\text{r}}$
antennas. The received signal of the $r$-th receive antenna at the
$k$-th symbol time is modeled as 
\begin{eqnarray*}
\boldsymbol{y}_{r,k} & = & \sum_{t=1}^{l_{\text{t}}}e^{j2\pi\boldsymbol{f}_{r,t}(k-1)}s_{t,k}\boldsymbol{h}_{r,t,k}+\boldsymbol{n}_{r,k},
\end{eqnarray*}
where $r=1,...,l_{\text{r}}$; $t=1,...,l_{\text{t}}$ is the transmit
antenna index; $k=1,...,n$ is the symbol time index; $\boldsymbol{h}_{r,t,k}\in\mathbb{C}$
is the channel coefficient from the $t$-th transmit antenna to the
$r$-th receive antenna; $\boldsymbol{n}_{r,k}\sim\mathcal{CN}\left(0,\sigma_{\boldsymbol{n}}^{2}\right)$,
$\sigma_{\boldsymbol{n}}^{2}=1$, $\forall r,k$, are i.i.d. circularly
symmetric complex Gaussian distributed with zero mean and unit variance;
$s_{t,k}\in\mathbb{C}$ is the pilot/training signal sent from the
$t$-th transmit antenna at time $k$; $\boldsymbol{f}_{r,t}=\bar{\boldsymbol{f}}_{r,t}t_{b}$
is the residual normalized carrier frequency offset (CFO) between
antennas $t$ and $r$, due to what is left from the coarse frequency
synchronization; $t_{b}$ is the symbol period; $\bar{\boldsymbol{f}}_{r,t}$
is the pre-normalized carrier frequency offset. In this paper, CFO
refers to $\boldsymbol{f}_{r,t}$. Collect the variables in vectors,
\begin{eqnarray*}
\underset{\vec{\boldsymbol{y}}}{\underbrace{\left[\underset{\vec{\boldsymbol{y}_{r}}}{\underbrace{\left[\boldsymbol{y}_{r,k}\right]_{k}}}\right]_{r}}} & = & \underset{\grave{\boldsymbol{X}}}{\underbrace{\left[\left[\left[e^{j2\pi\boldsymbol{f}_{r,t}(k-1)}s_{t,k}\right]_{1,t}\right]_{k,k}\right]_{r,r}}}\times\\
 &  & \underset{\vec{\boldsymbol{h}}}{\underbrace{\left[\underset{\vec{\boldsymbol{h}}_{r}}{\underbrace{\left[\left[\boldsymbol{h}_{r,t,k}\right]_{t}\right]_{k}}}\right]_{r}}}+\underset{\vec{\boldsymbol{n}}}{\underbrace{\left[\left[\boldsymbol{n}_{r,k}\right]_{k}\right]_{r}}},
\end{eqnarray*}
{} to obtain
\begin{eqnarray}
\vec{\boldsymbol{y}} & = & \grave{\boldsymbol{X}}\vec{\boldsymbol{h}}+\vec{\boldsymbol{n}}\in\mathbb{C}^{nl_{\text{r}}\times1}.\label{eq:tv_channel-model-fine-freq}
\end{eqnarray}

The spatially and time correlated channel state $\vec{\boldsymbol{h}}$
has distribution $\mathcal{CN}\left(\vec{\mu}_{\vec{\boldsymbol{h}}},\Sigma_{\vec{\boldsymbol{h}}}\right)$,
where $\vec{\mu}_{\vec{\boldsymbol{h}}}=\left[\left[\left[\mu_{\boldsymbol{h}_{r,t,k}}\right]_{t}\right]_{k}\right]_{r}\in\mathbb{C}^{l_{\text{t}}l_{\text{r}}n\times1}$
is the mean; and 
\begin{eqnarray}
\Sigma_{\vec{\boldsymbol{h}}} & = & \left[\left[\left[c_{\boldsymbol{h}_{r_{1},t_{1},k_{1}},\boldsymbol{h}_{r_{2},t_{2},k_{2}}}\right]_{t_{1},t_{2}}\right]_{k_{1},k_{2}}\right]_{r_{1},r_{2}}\in\mathbb{C}^{l_{\text{t}}l_{\text{r}}n\times l_{\text{t}}l_{\text{r}}n}\label{eq:tv_cov-h}
\end{eqnarray}
 is the covariance matrix of $\vec{\boldsymbol{h}}$ and $c_{\boldsymbol{h}_{r_{1},t_{1},k_{1}},\boldsymbol{h}_{r_{2},t_{2},k_{2}}}$
is the covariance between $\boldsymbol{h}_{r_{1},t_{1},k_{1}}$ and
$\boldsymbol{h}_{r_{2},t_{2},k_{2}}$. %
\begin{comment}
$\left[\text{E}\left[\vec{\boldsymbol{h}}_{r_{1}}\vec{\boldsymbol{h}}_{r_{2}}^{\dagger}\right]\right]_{r_{1}=1:l_{\text{r}},r_{2}=1:l_{\text{r}}}$
\end{comment}
{} %
\begin{comment}
We send pilot signal $\vec{s}_{t}\in\mathbb{C}^{1\times n}$ on the
$t$-th antenna. Collect the pilot signals in a matrix .
\end{comment}
The frequency offset $\vec{\boldsymbol{f}}=\left[\left[\boldsymbol{f}_{r,t}\right]_{t}\right]_{r}$
is approximated by a Gaussian distribution $\mathcal{N}(\mu_{\vec{\boldsymbol{f}}},\Sigma_{\vec{\boldsymbol{f}}})$
\cite{Tarokh_2008ITSP_BoundsAlgorithmsFrequencySynchronizationCollaborativeCommunicationSystems}.
Modern frequency sources of communication devices are typically stable.
In addition, after the coarse frequency synchronization, the residual
frequency offset is limited to a small range. Thus, the exponential
drop off of the Gaussian distribution is suitable. We have observed
in simulation that changing the distribution to others makes little
difference. The Gaussian distribution assumption is not used in the
derivation until (\ref{eq:tv_jointpdff}). The pilot signals have
average power $\rho=\frac{1}{n}\text{Tr}\left(S^{\dagger}S\right)$,
where $S=\left[s_{t,k}\right]_{k,t}\in\mathbb{C}^{n\times l_{\text{t}}}$.
We consider both the general case and the special case of orthogonal
pilots where $S^{\dagger}S=\frac{n\rho}{l_{\text{t}}}I_{l_{\text{t}}\times l_{\text{t}}}$,
which is a scaled identity matrix. %
{} %
\begin{comment}
the simplest case of i.i.d. channels $\text{E}\left[\vec{\boldsymbol{h}}_{r_{1}}\vec{\boldsymbol{h}}_{r_{2}}^{\dagger}\right]=\sigma_{\boldsymbol{h}}^{2}I\delta[r_{1}-r_{2}]$
and
\end{comment}
{} %
\begin{comment}
i.i.d. noise $\text{E}\left[\vec{\boldsymbol{n}}_{r_{1}}\vec{\boldsymbol{n}}_{r_{2}}^{\dagger}\right]=\sigma_{\boldsymbol{n}}^{2}I\delta[r_{1}-r_{2}]$,
whose variance of each element is normalized to $\sigma_{\boldsymbol{n}}^{2}=1$.
\end{comment}
{} %
\begin{comment}
It is the case of choice when we have little knowledge of the channel.
\end{comment}
{} %
\begin{comment}
To write the model in the familiar linear transformation form, we
define tall vectors
\end{comment}
\begin{comment}
\begin{eqnarray*}
\boldsymbol{Y}^{T} & = & \boldsymbol{H}\boldsymbol{X}^{T}+\boldsymbol{N}^{T}\\
=\left[\begin{array}{c}
\vec{\boldsymbol{y}}_{1}^{T}\\
\vdots\\
\vec{\boldsymbol{y}}_{l_{\text{r}}}^{T}
\end{array}\right] & = & \left[\begin{array}{c}
\vec{\boldsymbol{h}}_{1}^{T}\\
\vdots\\
\vec{\boldsymbol{h}}_{l_{\text{r}}}^{T}
\end{array}\right]\left[\begin{array}{c}
\vec{\boldsymbol{x}}_{1}^{T}\\
\vdots\\
\vec{\boldsymbol{x}}_{l_{\text{t}}}^{T}
\end{array}\right]+\left[\begin{array}{c}
\vec{\boldsymbol{n}}_{1}^{T}\\
\vdots\\
\vec{\boldsymbol{n}}_{l_{\text{r}}}^{T}
\end{array}\right]\\
 & = & \left[\begin{array}{c}
\vec{\boldsymbol{h}}_{1}^{T}\boldsymbol{X}^{T}+\vec{\boldsymbol{n}}_{1}^{T}\\
\vdots\\
\vec{\boldsymbol{h}}_{l_{\text{r}}}^{T}\boldsymbol{X}^{T}+\vec{\boldsymbol{n}}_{l_{\text{r}}}^{T}
\end{array}\right],
\end{eqnarray*}
\end{comment}
{} %
\begin{comment}
\begin{eqnarray*}
\vec{\boldsymbol{y}} & = & \left[\begin{array}{c}
\vec{\boldsymbol{y}}_{1}\\
\vdots\\
\vec{\boldsymbol{y}}_{l_{\text{r}}}
\end{array}\right],
\end{eqnarray*}
\begin{eqnarray*}
\vec{\boldsymbol{h}} & = & \left[\begin{array}{c}
\vec{\boldsymbol{h}}_{1}\\
\vdots\\
\vec{\boldsymbol{h}}_{l_{\text{r}}}
\end{array}\right],
\end{eqnarray*}
\begin{eqnarray*}
\vec{\boldsymbol{n}} & = & \left[\begin{array}{c}
\vec{\boldsymbol{n}}_{1}\\
\vdots\\
\vec{\boldsymbol{n}}_{l_{\text{r}}}
\end{array}\right],
\end{eqnarray*}
\end{comment}
{} %
\begin{comment}
$\vec{\boldsymbol{y}}=\left[\vec{\boldsymbol{y}}_{r}\right]_{r},$
$\vec{\boldsymbol{h}}=\left[\vec{\boldsymbol{h}}_{r}\right]_{r},$
$\vec{\boldsymbol{n}}=\left[\vec{\boldsymbol{n}}_{r}\right]_{r},$
and block matrix where $\boldsymbol{I}_{l_{\text{r}}}$ is a $l_{\text{r}}\times l_{\text{r}}$
identity matrix and $\otimes$ is the Kronecker product. With these
notation, the received signal can be written as
\end{comment}

Among the orthogonal pilots, we consider two typical representatives
\cite{Liu_2021TSP_MaximumPosterioriProbabilityMAPJointFineFrequencyOffsetChannelEstimationMIMOSystemsChannelsArbitraryCorrelation},
the periodic pilot and time-division (TD) pilot defined below. We
define \emph{Scrambled Periodic Pilot} as 
\begin{eqnarray}
S & = & \sqrt{\rho}\underset{C}{\underbrace{\left[c_{k}\right]_{k,k=1:n}}}\underset{\left[O\right]_{i=1:m}}{\underbrace{\left[I_{l_{\text{t}}}\right]_{i=1:m}O}},\label{eq:tv_scrambled-pilot-2}
\end{eqnarray}
where $I_{l_{\text{t}}}$ is an $l_{\text{t}}\times l_{\text{t}}$
identity matrix; $n=ml_{\text{t}}$ is assumed for $m\in\mathbb{Z}^{+}$.
It has a structure of scrambled periodic matrix $\left[O\right]_{i=1:m}=\left[\begin{array}{c}
O\\
O\\
\vdots
\end{array}\right]\in\mathbb{C}^{n\times l_{\text{t}}}$, which is a block matrix with $m$ copies of an unitary matrix $O\in\mathbb{C}^{l_{\text{t}}\times l_{\text{t}}}$
on top of each other. Matrix $O$ satisfies $O^{\dagger}O=OO^{\dagger}=I_{l_{\text{t}}}$.
The scrambling code is $\vec{c}=\left[c_{k}\right]_{k=1:n}\in\mathbb{C}^{n\times1}$,
where $|c_{k}|=1,\ \forall k$. Diagonal matrix $C$'s diagonal elements
are from $\vec{c}$. The choices of $\vec{c}$ and $O$ do not affect
performance. A simple example for $c_{k}=1$, $O=I_{l_{\text{t}}}$,
$m=2$, $l_{\text{t}}=3$ is
\begin{eqnarray*}
S & = & \sqrt{\rho}\left[\begin{array}{ccc}
1 & 0 & 0\\
0 & 1 & 0\\
0 & 0 & 1\\
1 & 0 & 0\\
0 & 1 & 0\\
0 & 0 & 1
\end{array}\right].
\end{eqnarray*}
\begin{comment}
Another example of this pilot structure is rows or columns of the
Hadamard matrix. 
\end{comment}
The freedom of choosing $C$ and $O$ offers flexibility for this
structure. %
\begin{comment}
For example, $O$ could be a Hadamard matrix or a Fourier transform
matrix $\left[e^{-j2\pi\frac{ik}{l_{\text{t}}}}\right]_{i=1:l_{\text{t}},k=1:l_{\text{t}}}$,
while $C$ could be a Gold or a Zadoff-Chu sequence \cite{Chu_1972ITIT_PolyphaseCodesGoodPeriodicCorrelationPropertiesCorresp}.
\end{comment}
{} Another typical pilot signal used in practice is the \emph{Time Division
(TD)Pilot} 
\begin{eqnarray}
S & = & \sqrt{\rho}\underset{C}{\underbrace{\left[c_{k}\right]_{k,k=1:n}}}\left[\vec{1}_{m}\right]_{i,i=1:l_{\text{t}}},\label{eq:tv_TDD-pilot-1}
\end{eqnarray}
where $n=ml_{\text{t}}$; only the first transmit antenna transmits
scrambled $m$ ones, followed by that only the second antenna transmits
$m$ scrambled ones, \emph{etc.}. Vector $\vec{1}_{m}$ has $m$ ones
on top of each other. Diagonal block matrix $\left[\vec{1}_{m}\right]_{i,i=1:l_{\text{t}}}=\left[\begin{array}{ccc}
\vec{1}_{m} & \vec{0} & \ldots\\
\vec{0} & \vec{1}_{m} & \cdots\\
\vdots & \vdots & \ddots
\end{array}\right]\in\mathbb{R}^{n\times l_{\text{t}}}$. A simple example for $c_{k}=1$, $m=2$, $l_{\text{t}}=3$ is
\begin{eqnarray*}
S & = & \sqrt{\rho}\left[\begin{array}{ccc}
1 & 0 & 0\\
1 & 0 & 0\\
0 & 1 & 0\\
0 & 1 & 0\\
0 & 0 & 1\\
0 & 0 & 1
\end{array}\right].
\end{eqnarray*}

We observe that the both $\left[I_{l_{\text{t}}}\right]_{i=1:m}$
and $\left[\vec{1}_{m}\right]_{i,i=1:l_{\text{t}}}$ have one $1$
per row and $m$ 1's per column. They represent two opposite ways
to arrange the rows and are useful in different scenarios and have
different performance. The periodic structure with $O$ is useful
when we do not want to switch on and off antennas. For the same amount
of signal energy, it only requires $\frac{1}{l_{\text{t}}}$ peak
power per antenna of the time division structure, because all antennas
are on all the time. The TD structure is useful when we need backward
compatibility to single antenna systems and when we can afford larger
peak power per antenna.

\section{The Optimization Problem and Solution\label{sec:tv_The-Optimization-Problem}}

To perform joint MAP estimation of channel and frequency offset, we
solve the following optimization problem. The following derivation
is the similar as the time invariant case.
\begin{problem}
\label{prob:tv_joint-fine-freq-channel} The problem of joint MAP
estimation of the fine frequency offset and the channel is
\begin{eqnarray}
 &  & (\hat{\vec{h}},\hat{\vec{f}})\nonumber \\
 & = & \arg\max_{\vec{h},\hat{\vec{f}}}f_{\vec{\boldsymbol{h}},\vec{\boldsymbol{f}},\vec{\boldsymbol{y}}}(\vec{h},\vec{f},\vec{y})\nonumber \\
 & = & \arg\max_{\vec{h},\vec{f}}f_{\vec{\boldsymbol{h}}|\vec{\boldsymbol{y}},\vec{\boldsymbol{f}}}(\vec{h}|\vec{y},\vec{f})f_{\vec{\boldsymbol{f}},\vec{\boldsymbol{y}}}(\vec{f},\vec{y})\nonumber \\
 & = & \arg\max_{\vec{f}}\left(\arg\max_{\vec{h}}f_{\vec{\boldsymbol{h}}|\vec{\boldsymbol{y}},\vec{\boldsymbol{f}}}(\vec{h}|\vec{y},\vec{f})\right)\nonumber \\
 &  & \times f_{\vec{\boldsymbol{y}}|\vec{\boldsymbol{f}}}(\vec{y}|\vec{f})f_{\vec{\boldsymbol{f}}}(\vec{f}).\label{eq:tv_MAP-h-f}
\end{eqnarray}
\end{problem}
\textbf{Solution:} The maximization over $\vec{f}$ and $\vec{h}$
appears coupled but are actually separable for $f_{r,t}=f_{r}$ or
$f_{r,t}=f$ cases, as shown in the following steps, similar to \cite{Liu_2021TSP_MaximumPosterioriProbabilityMAPJointFineFrequencyOffsetChannelEstimationMIMOSystemsChannelsArbitraryCorrelation}.
\begin{enumerate}
\item Perform the MAP estimation of the channel given a frequency offset
$\vec{f}$:
\begin{eqnarray}
\hat{\vec{h}}(\vec{y},\vec{f}) & = & \arg\max_{\vec{h}}f_{\vec{\boldsymbol{h}}|\vec{\boldsymbol{y}},\vec{\boldsymbol{f}}}(\vec{h}|\vec{y},\vec{f}).\label{eq:tv_max-h}
\end{eqnarray}
\item Substitute the above result in (\ref{eq:tv_MAP-h-f}) to estimate
the CFO using
\begin{eqnarray*}
\hat{\vec{f}} & = & \arg\max_{\vec{f}}f_{\vec{\boldsymbol{h}}|\vec{\boldsymbol{y}},\vec{\boldsymbol{f}}}(\hat{\vec{h}}(\vec{y},\vec{f})|\vec{y},\vec{f})\\
 &  & \times f_{\vec{\boldsymbol{y}}|\vec{\boldsymbol{f}}}(\vec{y}|\vec{f})f_{\vec{\boldsymbol{f}}}(\vec{f})\\
 & = & \arg\max_{\vec{f}}f_{\vec{\boldsymbol{y}}|\vec{\boldsymbol{f}}}(\vec{y}|\vec{f})f_{\vec{\boldsymbol{f}}}(\vec{f}),
\end{eqnarray*}
 where we show in Theorem \ref{thm:MAP-channel} that $f_{\vec{\boldsymbol{h}}|\vec{\boldsymbol{y}},\vec{\boldsymbol{f}}}(\hat{\vec{h}}(\vec{y},\vec{f})|\vec{y},\vec{f})$
is not a function of $\vec{f}$ if $f_{r,t}=f_{r}$ or $f_{r,t}=f$.
Therefore, the joint estimations of frequency offset and channel are
separable and we can solve an individual MAP estimation of $\vec{\boldsymbol{f}}$
with channel state distribution information. If needed, one can assume
that $\vec{\boldsymbol{f}}$ is uniform either over all real number
or over a small interval, or is Gaussian with infinite variance. Then,
the MAP estimation of $\vec{\boldsymbol{f}}$ can be converted to
the ML estimation,
\begin{eqnarray*}
\hat{\vec{f}} & = & \arg\max_{\vec{f}}f_{\vec{\boldsymbol{y}}|\vec{\boldsymbol{f}}}(\vec{y}|\vec{f})f_{\vec{\boldsymbol{f}}}(\vec{f})\\
 & = & \arg\max_{\vec{f}}f_{\vec{\boldsymbol{y}}|\vec{\boldsymbol{f}}}(\vec{y}|\vec{f}),
\end{eqnarray*}
over any value or in the small interval.
\item Finally, $\hat{\vec{h}}(\vec{y},\hat{\vec{f}})$ gives the solution
of the channel estimation.
\end{enumerate}

\subsection{MAP and ML Channel Estimation}

For the first step of the solution, we have the following theorem.
\begin{thm}
\label{thm:MAP-channel}The solution to (\ref{eq:tv_max-h}), the
MAP estimation of $\vec{\boldsymbol{h}}$ given $\vec{f}$ and $\vec{y}$,
is
\begin{eqnarray*}
\hat{\vec{h}}(\vec{y},\vec{f}) & = & \hat{\vec{h}}_{\text{MMSE}}(\vec{y},\vec{f}),
\end{eqnarray*}
which is given in (\ref{eq:h_MMSE}). And the density
\begin{eqnarray*}
f_{\vec{\boldsymbol{h}}|\vec{\boldsymbol{y}},\vec{\boldsymbol{f}}}(\hat{\vec{h}}(\vec{y},\vec{f})|\vec{y},\vec{f}) & = & \frac{1}{\det\left(\pi\Sigma_{\hat{\vec{\boldsymbol{h}}}_{\text{MMSE}}}\right)}
\end{eqnarray*}
 is not a function of $\vec{f}$ when $\boldsymbol{f}_{r,t}=\boldsymbol{f}_{r}$
or $\boldsymbol{f}_{r,t}=\boldsymbol{f}$, where $\Sigma_{\hat{\vec{\boldsymbol{h}}}_{\text{MMSE}}}$
is given in (\ref{eq:Sigma_hMMSE}).
\end{thm}
\begin{IEEEproof}
See Appendix \ref{sec:tv_Proof-of-MAP-channel}.
\end{IEEEproof}
\begin{rem}
Setting $\Sigma_{\vec{\boldsymbol{h}}}^{-1}=\boldsymbol{0}$ in the
above provides ML or least square channel estimation.
\end{rem}
\begin{rem}
\label{rem:tv_A-special-case}To gain insight of the BCRLB later,
we examine the special case of $\boldsymbol{f}_{r,t}=\boldsymbol{f}_{r}$
and spatially uncorrelated and zero mean channel. In such a case,
\begin{eqnarray*}
 &  & c_{\boldsymbol{h}_{r_{1},t_{1},k_{1}},\boldsymbol{h}_{r_{2},t_{2},k_{2}}}\\
 & = & \varrho_{\boldsymbol{h}}(k_{1},k_{2})\delta[r_{1}-r_{2}]\delta[t_{1}-t_{2}].
\end{eqnarray*}
Thus, 
\begin{eqnarray*}
\Sigma_{\vec{\boldsymbol{h}}} & = & \left[\left[\left[\varrho_{\boldsymbol{h}}(k_{1},k_{2})\right]_{t,t}\right]_{k_{1},k_{2}}\right]_{r,r}.
\end{eqnarray*}
 If the fading process is wide sense stationary, $\Sigma_{\vec{\boldsymbol{h}}}$
is block Toeplitz. So is $\Sigma_{\vec{\boldsymbol{h}}}^{-1}$. For
wide-sense stationary fading and periodic or TD pilot, it is possible
to calculate $A$ in terms of the power spectrum density of the fading
process. It is omitted here.%
{} 
\end{rem}

\subsection{MAP and ML Frequency Offset Estimation}

For the second step, we observe that conditioned on $\{\vec{\boldsymbol{f}}=\vec{f}\}$,
$\vec{\boldsymbol{y}}$ is a summation of Gaussian random variables
and has distribution $\mathcal{CN}\left(\vec{\mu}_{\vec{\boldsymbol{y}}|\vec{\boldsymbol{f}}}(\vec{f}),\Sigma_{\vec{\boldsymbol{y}}|\vec{\boldsymbol{f}}}(\vec{f})\right)$,
where 
\begin{eqnarray}
\Sigma_{\vec{\boldsymbol{y}}|\vec{\boldsymbol{f}}} & = & \grave{X}\Sigma_{\vec{\boldsymbol{h}}}\grave{X}^{\dagger}+I,\label{eq:tv_cov-y}
\end{eqnarray}
 according to \prettyref{eq:tv_channel-model-fine-freq}.%
\begin{comment}
\begin{eqnarray}
\Sigma_{\vec{\boldsymbol{y}}|\vec{\boldsymbol{f}}} & = & \grave{X}\Sigma_{\vec{\boldsymbol{h}}}\grave{X}^{\dagger}+I\nonumber \\
 & = & ?\nonumber \\
 & = & \sigma_{\boldsymbol{h}}^{2}\left[\begin{array}{ccc}
XX^{\dagger} & \mathbf{0} & \mathbf{0}\\
\mathbf{0} & \ddots & \mathbf{0}\\
\mathbf{0} & \mathbf{0} & XX^{\dagger}
\end{array}\right]+I\nonumber \\
 & = & \sigma_{\boldsymbol{h}}^{2}I_{l_{\text{r}}}\otimes(FSS^{\dagger}F^{\dagger})+I.\label{eq:tv_cov-y-1}
\end{eqnarray}
\end{comment}
{} Using identity $\det(I+AB)=\det(I+BA)$, we obtain 
\begin{eqnarray}
 &  & \det(\pi\Sigma_{\vec{\boldsymbol{y}}|\vec{\boldsymbol{f}}})\nonumber \\
 & = & \left(\pi\right)^{nl_{\text{r}}}\det\left(I+\Sigma_{\vec{\boldsymbol{h}}}\grave{X}^{\dagger}\grave{X}\right)\nonumber \\
 & = & \begin{cases}
\left(\pi\right)^{nl_{\text{r}}}\det\left(I+\Sigma_{\vec{\boldsymbol{h}}}\grave{S}^{\dagger}\grave{S}\right) & \boldsymbol{f}_{r,t_{2}}=\boldsymbol{f}_{r,t_{1}},\ \forall t_{1},t_{2}\\
\left(\pi\right)^{nl_{\text{r}}}\det\left(I+\Sigma_{\vec{\boldsymbol{h}}}\grave{X}^{\dagger}\grave{X}\right) & \text{else}
\end{cases},\label{eq:tv_det-not-depend-on-f}
\end{eqnarray}
which is not a function of $\vec{f}$ when the frequencies may deffer
by the receive antenna indexes. We have the following theorem.
\begin{thm}
\label{thm:tv_MAP-f}For Gaussian distributed random channel $\vec{\boldsymbol{h}}$
and $\boldsymbol{f}_{r,t}=\boldsymbol{f}_{r}$, $\vec{\boldsymbol{f}}=\left[\boldsymbol{f}_{r}\right]_{r}$,
the MAP frequency offset estimate is 
\begin{eqnarray}
\hat{\vec{f}} & = & \arg\max_{\vec{f}}f_{\vec{\boldsymbol{y}}|\vec{\boldsymbol{f}}}(\vec{y}|\vec{f})f_{\vec{\boldsymbol{f}}}(\vec{f})\nonumber \\
 & = & \arg\max_{\vec{f}}\frac{1}{\det(\pi\Sigma_{\vec{\boldsymbol{y}}|\vec{\boldsymbol{f}}})}e^{-\left(\vec{y}-\vec{\mu}_{\vec{\boldsymbol{y}}|\vec{\boldsymbol{f}}}\right)^{\dagger}\Sigma_{\vec{\boldsymbol{y}}|\vec{\boldsymbol{f}}}^{-1}\left(\vec{y}-\vec{\mu}_{\vec{\boldsymbol{y}}|\vec{\boldsymbol{f}}}\right)}\nonumber \\
 &  & \times\frac{1}{\sqrt{2\pi\Sigma_{\vec{\boldsymbol{f}}}}}e^{-\frac{1}{2}\left(\vec{f}-\vec{\mu}_{\vec{\boldsymbol{f}}}\right)^{\dagger}\Sigma_{\vec{\boldsymbol{f}}}^{-1}\left(\vec{f}-\vec{\mu}_{\vec{\boldsymbol{f}}}\right)}\label{eq:tv_jointpdff}\\
 & = & \arg\max_{\vec{f}}g(\vec{y},\vec{f}),\label{eq:tv_max-g}
\end{eqnarray}
 where 
\begin{eqnarray}
 &  & g(\vec{y},\vec{f})\nonumber \\
 & \triangleq & 2\Re\left[\left\langle \grave{X}^{\dagger}\vec{y},\vec{b}\right\rangle \right]+\left(\grave{X}^{\dagger}\vec{y}\right)^{\dagger}A\left(\grave{X}^{\dagger}\vec{y}\right)\nonumber \\
 &  & -\frac{1}{2}\vec{f}^{\dagger}\Sigma_{\vec{\boldsymbol{f}}}^{-1}\vec{f}+\Re\left[\vec{\mu}_{\vec{\boldsymbol{f}}}^{\dagger}\Sigma_{\vec{\boldsymbol{f}}}^{-1}\vec{f}\right];\label{eq:tv_gyf}
\end{eqnarray}
\begin{comment}
component format for programming:
\begin{eqnarray*}
 &  & g(\vec{y},\vec{f})\\
 & \triangleq & 2\Re\left[\left\langle \grave{X}^{\dagger}\vec{y},\vec{b}\right\rangle \right]+\left(\grave{X}^{\dagger}\vec{y}\right)^{\dagger}A\left(\grave{X}^{\dagger}\vec{y}\right)\\
 &  & -\frac{1}{2}\vec{f}^{\dagger}\Sigma_{\vec{\boldsymbol{f}}}^{-1}\vec{f}+\Re\left[\vec{\mu}_{\vec{\boldsymbol{f}}}^{\dagger}\Sigma_{\vec{\boldsymbol{f}}}^{-1}\vec{f}\right]\\
 & = & 2\Re\left[\sum_{r,t,k}e^{j2\pi f_{r}(k-1)}s_{t,k}y_{r,k}^{*}b_{r,t,k}\right]+\\
 &  & \sum_{r_{1},t_{1},k_{1},r_{2},t_{2},k_{2}}e^{j2\pi\left(f_{r_{1}}(k_{1}-1)-f_{r_{2}}(k_{2}-1)\right)}a_{r_{1},t_{1},k_{1},r_{2},t_{2},k_{2}}\times\\
 &  & s_{t_{1},k_{1}}s_{t_{2},k_{2}}^{*}y_{r_{2},k_{2}}y_{r_{1},k_{1}}^{*}+\\
 &  & \sum_{r}\left(-\frac{1}{2}f_{r}^{2}\sigma_{\boldsymbol{f}_{r}}^{-2}+\Re\left[\mu_{\boldsymbol{f}_{r}}^{*}\sigma_{\boldsymbol{f}_{r}}^{-2}f_{r}\right]\right)
\end{eqnarray*}

matrix 
\begin{eqnarray*}
 &  & g(\vec{y},\vec{f})\\
 & \triangleq & 2\Re\left[\left\langle \grave{X}^{\dagger}\vec{y},\vec{\mu}_{\vec{\boldsymbol{h}}}\right\rangle \right]\\
 &  & +\left(\grave{X}^{\dagger}\vec{y}-\frac{n\rho}{l_{\text{t}}}\vec{\mu}_{\vec{\boldsymbol{h}}}\right)^{\dagger}A\left(\grave{X}^{\dagger}\vec{y}-\frac{n\rho}{l_{\text{t}}}\vec{\mu}_{\vec{\boldsymbol{h}}}\right)\\
 &  & -\sigma_{\vec{\boldsymbol{f}}}^{-2}\left|\vec{f}-\vec{\mu}_{\vec{\boldsymbol{f}}}\right|^{2}.
\end{eqnarray*}
\end{comment}
{} $A$ is given in \prettyref{eq:tv_def-A} and $\vec{b}$ is given
in \prettyref{eq:tv_def-b}, which are not functions of $\vec{f}$;
$\grave{X}$ is a function of $\vec{f}$. The ML estimator is obtained
by setting $\Sigma_{\vec{\boldsymbol{f}}}^{-1}=\boldsymbol{0}$ in
\prettyref{eq:tv_gyf}.
\end{thm}
The proof is given in Appendix \ref{sec:tv_Proof-of-MAP}. When $f_{\vec{\boldsymbol{f}}}$
has a uniform distribution, the MAP estimator becomes the ML estimator,
corresponding to $\Sigma_{\vec{\boldsymbol{f}}}^{-1}=\boldsymbol{0}$.
For independent $\vec{\boldsymbol{f}}$, $\Sigma_{\vec{\boldsymbol{f}}}=\left[\sigma_{\boldsymbol{f}_{r}}^{2}\right]_{r,r}$. 

To understand $g$, we rewrite it as 
\begin{eqnarray*}
 &  & g(\vec{y},\vec{f})\\
 & = & 2\Re\left[\left\langle \grave{X}^{\dagger}\vec{y},\vec{b}\right\rangle \right]+\left(\grave{X}^{\dagger}\vec{y}\right)^{\dagger}A\left(\grave{X}^{\dagger}\left(\vec{y}-\grave{X}\vec{\mu}_{\vec{\boldsymbol{h}}}+\grave{X}\vec{\mu}_{\vec{\boldsymbol{h}}}\right)\right)\\
 &  & -\frac{1}{2}\vec{f}^{\dagger}\Sigma_{\vec{\boldsymbol{f}}}^{-1}\vec{f}+\Re\left[\vec{\mu}_{\vec{\boldsymbol{f}}}^{\dagger}\Sigma_{\vec{\boldsymbol{f}}}^{-1}\vec{f}\right]\\
 & = & 2\Re\left[\left\langle \grave{X}^{\dagger}\vec{y},\vec{b}\right\rangle \right]+\left(\grave{X}^{\dagger}\vec{y}\right)^{\dagger}\left(\hat{\vec{h}}_{\text{MMSE}}(\vec{y},\vec{f})-\vec{\mu}_{\vec{\boldsymbol{h}}}\right)\\
 &  & +\left(\grave{X}^{\dagger}\vec{y}\right)^{\dagger}A\left(\grave{X}^{\dagger}\grave{X}\vec{\mu}_{\vec{\boldsymbol{h}}}\right)\\
 &  & -\frac{1}{2}\vec{f}^{\dagger}\Sigma_{\vec{\boldsymbol{f}}}^{-1}\vec{f}+\Re\left[\vec{\mu}_{\vec{\boldsymbol{f}}}^{\dagger}\Sigma_{\vec{\boldsymbol{f}}}^{-1}\vec{f}\right]\\
 & = & 2\Re\left[\left\langle \grave{X}^{\dagger}\vec{y},\vec{b}\right\rangle \right]+\left(\grave{X}^{\dagger}\vec{y}\right)^{\dagger}\left(\hat{\vec{h}}_{\text{MMSE}}(\vec{y},\vec{f})-\vec{\mu}_{\vec{\boldsymbol{h}}}\right)\\
 &  & +\left(\grave{X}^{\dagger}\vec{y}\right)^{\dagger}\left(\vec{\mu}_{\vec{\boldsymbol{h}}}-\vec{b}\right)\\
 &  & -\frac{1}{2}\vec{f}^{\dagger}\Sigma_{\vec{\boldsymbol{f}}}^{-1}\vec{f}+\Re\left[\vec{\mu}_{\vec{\boldsymbol{f}}}^{\dagger}\Sigma_{\vec{\boldsymbol{f}}}^{-1}\vec{f}\right]\\
 & = & 2\Re\left[\left\langle \grave{X}^{\dagger}\vec{y},\vec{b}\right\rangle \right]+\vec{y}^{\dagger}\left(\grave{X}\hat{\vec{h}}_{\text{MMSE}}(\vec{y},\vec{f})\right)\\
 &  & -\left(\grave{X}^{\dagger}\vec{y}\right)^{\dagger}\vec{b}\\
 &  & -\frac{1}{2}\vec{f}^{\dagger}\Sigma_{\vec{\boldsymbol{f}}}^{-1}\vec{f}+\Re\left[\vec{\mu}_{\vec{\boldsymbol{f}}}^{\dagger}\Sigma_{\vec{\boldsymbol{f}}}^{-1}\vec{f}\right]\\
 & = & \left(\vec{y}^{\dagger}\grave{X}\vec{b}\right)^{\dagger}+\vec{y}^{\dagger}\left(\grave{X}\hat{\vec{h}}_{\text{MMSE}}(\vec{y},\vec{f})\right)\\
 &  & -\frac{1}{2}\vec{f}^{\dagger}\Sigma_{\vec{\boldsymbol{f}}}^{-1}\vec{f}+\Re\left[\vec{\mu}_{\vec{\boldsymbol{f}}}^{\dagger}\Sigma_{\vec{\boldsymbol{f}}}^{-1}\vec{f}\right],
\end{eqnarray*}
which is in terms of the MMSE estimate of the channel. The correlation
needs to be maximized.

The above proves the following theorem on the separable solution.

\begin{thm}
The joint fine frequency offset and channel estimation problem \ref{prob:tv_joint-fine-freq-channel}
can be decomposed into two separable optimization problems when $\boldsymbol{f}_{r,t}=\boldsymbol{f}_{r}$,
and $\vec{\boldsymbol{f}}=\left[\boldsymbol{f}_{r}\right]_{r}$:
\begin{enumerate}
\item The MAP estimation of $\vec{\boldsymbol{f}}$ is 
\begin{eqnarray*}
\hat{\vec{f}} & = & \arg\max_{\vec{f}}f_{\vec{\boldsymbol{y}}|\vec{\boldsymbol{f}}}(\vec{y}|\vec{f})f_{\vec{\boldsymbol{f}}}(\vec{f})=\arg\max_{\vec{f}}g(\vec{y},\vec{f}).
\end{eqnarray*}
Setting $f_{\vec{\boldsymbol{f}}}(\vec{f})$ as a constant, or making
$\Sigma_{\vec{\boldsymbol{f}}}^{-1}=\boldsymbol{0}$, it reduces to
the ML estimation of $\vec{\boldsymbol{f}}$. %
\begin{comment}
\begin{eqnarray}
\hat{\vec{f}} & = & \arg\max_{\vec{f}}f_{\vec{\boldsymbol{y}}|\vec{\boldsymbol{f}}}(\vec{y}|\vec{f})\nonumber \\
 & = & \arg\max_{\vec{f}}\underset{g(\vec{y},\vec{f})}{\underbrace{\sum_{r}\vec{y}_{r}^{\dagger}FSS^{\dagger}F^{\dagger}\vec{y}_{r}}}\label{eq:tv_est-freq-1}
\end{eqnarray}
\end{comment}
\item MAP or MMSE estimation of $\vec{\boldsymbol{h}}$ given the above
$\hat{\vec{f}}$ is%
\begin{comment}
\begin{enumerate}
\item 
\begin{eqnarray*}
\hat{\vec{h}}(\vec{y},\hat{\vec{f}}) & = & \arg\max_{\vec{h}}f_{\vec{\boldsymbol{h}}|\vec{\boldsymbol{y}},\vec{\boldsymbol{f}}}(\vec{h}|\vec{y},\hat{\vec{f}})=\hat{\vec{h}}_{\text{MMSE}}(\vec{y},\vec{f})\\
 & = & \left(\frac{n\rho}{l_{\text{t}}}+\frac{1}{\sigma_{\boldsymbol{h}}^{2}}\right)^{-1}\left[S^{\dagger}F(\hat{\vec{f}})^{\dagger}\vec{y}_{r}\right]_{r}.
\end{eqnarray*}
\end{enumerate}
\end{comment}
\begin{eqnarray*}
\hat{\vec{h}}(\vec{y},\hat{\vec{f}}) & = & \arg\max_{\vec{h}}f_{\vec{\boldsymbol{h}}|\vec{\boldsymbol{y}},\vec{\boldsymbol{f}}}(\vec{h}|\vec{y},\hat{\vec{f}})=\hat{\vec{h}}_{\text{MMSE}}(\vec{y},\hat{\vec{f}}).
\end{eqnarray*}
\end{enumerate}
\end{thm}
\begin{rem}
Setting $\Sigma_{\vec{\boldsymbol{h}}}^{-1}=\boldsymbol{0}$ in the
above provides frequency estimation without prior knowledge on channel
as in ML estimation.
\end{rem}
The MMSE estimation of the channel is straightforward. We focus on
the frequency offset estimation algorithms.

\section{Fine Frequency Offset Estimation Algorithms\label{sec:tv_Fine-Frequency-Synchronization}}

We design a low complexity algorithm for frequency offset estimation
for the case of $\boldsymbol{f}_{r,t}=\boldsymbol{f}_{r}$.

\subsection{Stationary Point Condition}

The intuitive meaning of the frequency offset estimation (\ref{eq:tv_max-g})
is to find $\vec{f}$ to de-rotate $\vec{y}$ so that its energy projected
to the signal space is maximized \cite{Liu_2021TSP_MaximumPosterioriProbabilityMAPJointFineFrequencyOffsetChannelEstimationMIMOSystemsChannelsArbitraryCorrelation}.
We may do so by solving stationary point condition $\frac{\partial g(\vec{y},\vec{f})}{\partial\vec{f}}=\vec{0}$.
It is summarized in the following theorem, with the special case of
$\boldsymbol{f}_{r}=\boldsymbol{f}$ as well.
\begin{thm}
\label{thm:tv_dgdf-0}For independent $\vec{\boldsymbol{f}}$, $\Sigma_{\vec{\boldsymbol{f}}}=\left[\sigma_{\boldsymbol{f}_{r}}^{2}\right]_{r,r}$,
the optimal solution $\vec{f}=\left[f_{r}\right]_{r}$ to the MAP
estimation problem satisfies
\begin{eqnarray}
\vec{0} & = & \frac{\partial g(\vec{y},\vec{f})}{\partial\vec{f}}\nonumber \\
 & = & -4\pi\Im[\nonumber \\
 &  & \sum_{k=1}^{n-1}e^{j2\pi f_{r}k}k\sum_{t}s_{t,k+1}y_{r,k+1}^{*}b_{r,t,k+1}+\nonumber \\
 &  & \sum_{k_{1}}(k_{1}-1)(\nonumber \\
 &  & \sum_{r_{2}:r<r_{2}}e^{j2\pi(f_{r}-f_{r_{2}})(k_{1}-1)}\eta_{r,k_{1},r_{2},k_{1}}-\nonumber \\
 &  & \sum_{r_{1}:r_{1}<r}e^{j2\pi(f_{r_{1}}-f_{r})(k_{1}-1)}\eta_{r_{1},k_{1},r,k_{1}})+\nonumber \\
 &  & \sum_{k_{1},k_{2}:k_{1}<k_{2}}(\sum_{r_{2}}(k_{1}-1)\nonumber \\
 &  & e^{j2\pi\left(f_{r}(k_{1}-1)-f_{r_{2}}(k_{2}-1)\right)}\eta_{r,k_{1},r_{2},k_{2}}-\nonumber \\
 &  & \sum_{r_{1}}(k_{2}-1)\nonumber \\
 &  & e^{j2\pi\left(f_{r_{1}}(k_{1}-1)-f_{r}(k_{2}-1)\right)}\eta_{r_{1},k_{1},r,k_{2}})]\nonumber \\
 &  & -\sigma_{\boldsymbol{f}_{r}}^{2}(f_{r}-\mu_{\boldsymbol{f}_{r}}).\label{eq:tv_dg-df-fr}
\end{eqnarray}
 If $\boldsymbol{f}_{r}=\boldsymbol{f}$, the optimal $f$ satisfies
\begin{eqnarray}
0 & = & \frac{\partial g(\vec{y},f)}{\partial f}\nonumber \\
 & = & -4\pi\Im\left[\sum_{k=1}^{n-1}e^{j2\pi fk}kz_{k}\right]\nonumber \\
 &  & -\sigma_{\boldsymbol{f}}^{-2}\left(f-\vec{\mu}_{\boldsymbol{f}}\right),\label{eq:tv_dg-df-f}
\end{eqnarray}
where $r_{k}>0$, %
\begin{comment}
\begin{eqnarray*}
\Sigma_{\vec{\boldsymbol{f}}}^{-1} & = & \left[\varsigma_{r_{1},r_{2}}\right]_{r_{1},r_{2}},
\end{eqnarray*}
\end{comment}
\begin{eqnarray}
z_{k} & \triangleq & r_{k}e^{-j\theta_{k}}\nonumber \\
 & \triangleq & \sum_{r,t}s_{t,k+1}y_{r,k+1}^{*}b_{r,t,k+1}+\nonumber \\
 &  & \sum_{k_{1}=k+1}^{n}\sum_{r_{1},t_{1},r_{2},t_{2}}a_{r_{1},t_{1},k_{1},r_{2},t_{2},k_{1}-k}\times\nonumber \\
 &  & s_{t_{1},k_{1}}s_{t_{2},k_{1}-k}^{*}y_{r_{2},k_{1}-k}y_{r_{1},k_{1}}^{*}.\label{eq:tv_rtheta-general}
\end{eqnarray}
\end{thm}
%
\begin{comment}
\begin{itemize}
\item {*}Assume zero mean iid channel produces better ML metric. It means
the ML metric is not wrong. It is the sin(x)=x and/or derivative is
wrong. It is more likely that sin(x)=x is wrong. sin(x)=x may not
apply to the term involves b.
\item {*}So, we need better zero noise analysis to find when sin(x)=x. If
not, a different method is needed.
\item {*} If we use second order approximation of exp(f), then the quadratic
equation for the exponent means that f has Gaussian distribution.
Then, making derivative zero produces conditional mean and the second
order derivative produces the 1/MMSE.
\begin{itemize}
\item Thus, if exp(f) can be approximated by the second order Taylor series,
MAP is close to MMSE.
\end{itemize}
\end{itemize}
\end{comment}

It is proved in Appendix \ref{sec:tv_Cal-dgdf}. To understand (\ref{eq:tv_dg-df-f}),
we examine 
\begin{eqnarray*}
 &  & \mathbb{E}\left[\left.\boldsymbol{y}_{r_{2},k_{1}-k}\boldsymbol{y}_{r_{1},k_{1}}^{*}\right|\boldsymbol{f}\right]\\
 & = & \mathbb{E}\left[\left(\sum_{t_{2}'=1}^{l_{\text{t}}}e^{j2\pi\boldsymbol{f}(k_{1}-k-1)}s_{t_{2}',k_{1}-k}\boldsymbol{h}_{r_{2},t_{2}',k_{1}-k}+\boldsymbol{n}_{r_{2},k_{1}-k}\right)\right.\\
 &  & \left.\left.\left(\sum_{t_{1}'=1}^{l_{\text{t}}}e^{-j2\pi\boldsymbol{f}(k_{1}-1)}s_{t_{1}',k_{1}}^{*}\boldsymbol{h}_{r_{1},t_{1}',k_{1}}^{*}+\boldsymbol{n}_{r_{1},k_{1}}^{*}\right)\right|\boldsymbol{f}\right]\\
 & = & e^{-j2\pi\boldsymbol{f}k}\sum_{t_{1}',t_{2}'}s_{t_{2}',k_{1}-k}s_{t_{1}',k_{1}}^{*}\\
 &  & \left(c_{\boldsymbol{h}_{r_{1},t_{1}',k_{1}},\boldsymbol{h}_{r_{2},t_{2}',k_{1}-k}}^{*}+\mu_{\boldsymbol{h}_{r_{2},t_{2}',k_{1}-k}}\mu_{\boldsymbol{h}_{r_{1},t_{1}',k_{1}}}^{*}\right),
\end{eqnarray*}
\begin{eqnarray*}
 &  & \mathbb{E}\left[\left.\sum_{k=1}^{n-1}e^{j2\pi fk}k\boldsymbol{z}_{k}\right|\boldsymbol{f}\right]\\
 & = & \sum_{k=1}^{n-1}e^{j2\pi(f-\boldsymbol{f})k}k\left(\sum_{r,t}s_{t,k+1}\sum_{t'}s_{t',k+1}^{*}\mu_{\boldsymbol{h}_{r,t',k+1}}^{*}b_{r,t,k+1}\right.\\
 &  & +\sum_{k_{1}=k+1}^{n}\sum_{r_{1},t_{1},r_{2},t_{2}}a_{r_{1},t_{1},k_{1},r_{2},t_{2},k_{1}-k}\times\\
 &  & s_{t_{1},k_{1}}s_{t_{2},k_{1}-k}^{*}\sum_{t_{1}',t_{2}'}s_{t_{2}',k_{1}-k}s_{t_{1}',k_{1}}^{*}\\
 &  & \left.\left(c_{\boldsymbol{h}_{r_{1},t_{1}',k_{1}},\boldsymbol{h}_{r_{2},t_{2}',k_{1}-k}}^{*}+\mu_{\boldsymbol{h}_{r_{2},t_{2}',k_{1}-k}}\mu_{\boldsymbol{h}_{r_{1},t_{1}',k_{1}}}^{*}\right)\right),
\end{eqnarray*}
which is similar to the CRLB derived later.%
\begin{comment}
\begin{eqnarray}
\beta & = & 8\pi^{2}\Re\left[\sum_{k=1}^{n-1}k^{2}\times\right.\nonumber \\
 &  & \left(\sum_{r,t}\sum_{t'}s_{t,k+1}s_{t',k+1}^{*}\mu_{\boldsymbol{h}_{r,t',k+1}}^{*}b_{r,t,k+1}+\right.\nonumber \\
 &  & \sum_{k_{1}=k+1}^{n}\sum_{r_{1},t_{1},r_{2},t_{2}}a_{r_{1},t_{1},k_{1},r_{2},t_{2},k_{1}-k}\times\nonumber \\
 &  & s_{t_{1},k_{1}}s_{t_{2},k_{1}-k}^{*}\sum_{t_{2}'}s_{t_{2}',k_{1}-k}\sum_{t_{1}'}s_{t_{1}',k_{1}}^{*}\times\nonumber \\
 &  & \left.\left.\left(c_{\boldsymbol{h}_{r_{1},t_{1}',k_{1}},\boldsymbol{h}_{r_{2},t_{2}',k_{1}-k}}^{*}+\mu_{\boldsymbol{h}_{r_{2},t_{2}',k_{1}-k}}\mu_{\boldsymbol{h}_{r_{1},t_{1}',k_{1}}}^{*}\right)\right)\right];\label{eq:tv_-E-general-1}
\end{eqnarray}
\end{comment}

\subsection{Universal Algorithm for the Case of $\boldsymbol{f}_{r}=\boldsymbol{f}$}

We take the same approach as in the time invariant case \cite{Liu_2021TSP_MaximumPosterioriProbabilityMAPJointFineFrequencyOffsetChannelEstimationMIMOSystemsChannelsArbitraryCorrelation}
to design an algorithm to solve the stationary point condition (\ref{eq:tv_dg-df-f}).
Let $f=f_{0}+f_{e}$. If we can find $f_{0}$ such that $f_{e}$ is
small, then the asymptotically accurate approximation, 
\begin{eqnarray}
e^{j2\pi fk} & = & e^{j2\pi(f_{0}+f_{e})k}\nonumber \\
 & = & e^{j2\pi f_{0}k}e^{j2\pi f_{e}k}\nonumber \\
 & \doteq & e^{j2\pi f_{0}k}\left(1+j2\pi f_{e}k\right),\label{eq:tv_expfe-tylor}
\end{eqnarray}
can be used in (\ref{eq:tv_dg-df-f}) to solve a linear equation of
$f_{e}$ to obtain 
\begin{eqnarray}
f_{e} & \doteq & \frac{-\frac{1}{2\pi}\Im\left[\sum_{k=1}^{n-1}ke^{j2\pi f_{0}k}z_{k}\right]+\frac{\sigma_{\vec{\boldsymbol{f}}}^{-2}}{8\pi^{2}}(\vec{\mu}_{\boldsymbol{f}}-f_{0})}{\Re\left[\sum_{k=1}^{n-1}k^{2}e^{j2\pi f_{0}k}z_{k}\right]+\frac{\sigma_{\boldsymbol{f}}^{-2}}{8\pi^{2}}}.\label{eq:tv_fe}
\end{eqnarray}

This approach does not rely on the assumption that $\frac{\theta_{k}+m_{k}2\pi}{2\pi k}$,
which depends on noise and other parameters, approaches $f$ and thus
\begin{eqnarray*}
\Im\left[e^{j2\pi fk}e^{-j\theta_{k}}\right] & = & \sin\left(2\pi\left(f-\frac{\theta_{k}+m_{k}2\pi}{2\pi k}\right)k\right)\\
 & \doteq & 2\pi\left(f-\frac{\theta_{k}+m_{k}2\pi}{2\pi k}\right)k,
\end{eqnarray*}
where $m_{k}$ is from phase unwrapping. This assumption used in the
past literature does not hold for correlated and nonzero mean channels
for some pilots even for the zero noise case, resulting in error floor.
In contrast, the approximation in (\ref{eq:tv_expfe-tylor}) is asymptotically
accurate, regardless the behavior of $\theta_{k}$, as long as $f_{0}$
is close to $\boldsymbol{f}$, which can be achieved by searching
$f_{0}$ on a coarse grid as small as $4n$ points in region $[-0.5,0.5)$. 

For each $f_{0}$ in the grid, we can calculate $f_{e}$. Choose the
$f_{0}+f_{e}$ that maximizes the MAP metric $g(\vec{y},f_{0}+f_{e})$.
Then, we can further improve the accuracy by replacing $f_{0}$ with
$f_{0}+f_{e}$ and finding a smaller $f_{e}$ iteratively. Consequently,
we obtain Algorithm \ref{alg:estimate-freq-more-general} that is
different from all past literature and is applicable in all situations
without error floor. In the algorithm, no more than a few rounds of
``while'' loop execution is needed before the convergence with grid
size $4n$ and $\epsilon=10^{-10}$ or smaller. No performance gain
is observed with finer grid. The maximum number of iteration of 10
is used to avoid limit cycles for bad $f_{0}$ choice. Since the fine
CFO is normally less than 10\% of the symbol rate, the search grid
can be further reduced. For each search step, the calculation in (\ref{eq:tv_fe})
does not need phase unwrapping and is in a closed form. Therefore,
the overall complexity is low.

\begin{algorithm}[h]
\caption{\label{alg:estimate-freq-more-general}Universal Frequency Offset
Estimation}

\begin{enumerate}
\item \textbf{Input: }Matched filter output $y_{r,k}$, $r=1,...,l_{\text{r}}$,
$k=1,...,n$.
\item Calculate $z_{k}$, $k=1,...,n-1$, by \prettyref{eq:tv_rtheta-general}
\item for $f_{0}$ on a grid within $[-0.5,0.5)$:
\begin{enumerate}
\item Calculate $f_{e}$ by (\ref{eq:tv_fe})
\item Record $f_{\delta}=f_{e}+f_{0}$ that achieves the largest MAP metric
$g(\vec{y},f_{\delta})$ of (\ref{eq:tv_gyf})
\end{enumerate}
\item $f_{0}=f_{\delta}$
\item while $f_{e}>\epsilon$ and the number of iterations < 10:
\begin{enumerate}
\item Calculate $f_{e}$ by (\ref{eq:tv_fe})
\item $f_{0}=f_{e}+f_{0}$
\end{enumerate}
\item \textbf{Output:} $\hat{f}_{\delta}=f_{0}$.
\end{enumerate}
\end{algorithm}

\begin{rem}
An alternative way to use $\vec{\mu}_{\vec{\boldsymbol{f}}}$ is to
de-rotate the received signals by $e^{-j2\pi\vec{\mu}_{\vec{\boldsymbol{f}}}k}$
and then estimate the frequency offset by setting $\vec{\mu}_{\vec{\boldsymbol{f}}}=0$.
The advantage is to increase the acquisition range of $|\vec{\boldsymbol{f}}|$
to the range of $|\vec{\boldsymbol{f}}-\vec{\mu}_{\vec{\boldsymbol{f}}}|$. 
\end{rem}
%
\begin{comment}
If we want to use a closed loop approach like phase lock loop, based
on (\ref{eq:tv_dg-df-general}), we can use 
\begin{eqnarray*}
e & = & -\gamma\sum_{k=1}^{n-1}kr_{k}\sin\left(2\pi k\left(\vec{f}-\frac{\theta_{k}}{2\pi k}\right)\right)
\end{eqnarray*}
 as the feedback error, where $\gamma$ is an appropriate step size.
This is equivalent to the smoothing filter approach when the filter
has feedback loops.
\end{comment}

\begin{rem}
Our MAP estimation of channel and frequency offset can be employed
to deal with time varying CFO cases by updating the prior distribution
of CFO for the next packet of data. See \cite{Liu_2021TSP_MaximumPosterioriProbabilityMAPJointFineFrequencyOffsetChannelEstimationMIMOSystemsChannelsArbitraryCorrelation}
for a case study. %
\begin{comment}
For example, if $\vec{\boldsymbol{f}}$ is time varying from packet
to packet, we can use current estimate, the estimation error variance,
to be calculated from the Bayesian Cramer-Rao lower bound in Section
\prettyref{sec:tv_Performance-Analysis}, and the correlation between
the current and the next frequency offset to calculate the prior distribution
of the next frequency offset. The prior distribution then is used
in the MAP estimation of the next packet/frame's frequency offset.
\end{comment}
\end{rem}

\subsection{Universal Algorithm for the Case of Different $\boldsymbol{f}_{r}$ }

The universal algorithm can be extended to the case of different $\boldsymbol{f}_{r}$
for different receive antennas. Each $f_{r}$ in (\ref{eq:tv_dg-df-fr})
can be replaced by $f_{r,0}+f_{r,e}$. Then, the same approximation
of (\ref{eq:tv_expfe-tylor}) can be applied to obtain linear equations
of $f_{r,e}$, which is easy to solve. To avoid $O(n^{l_{\text{r}}})$
complexity grid search of $f_{r,0}$'s, we can ignore the receive
antenna correlation and search and solve for $f_{r,0}$ using the
universal algorithm for each $r$. Then, these $f_{r,0}$'s can be
used together to refine $f_{r,e}$'s jointly using the linear equations.
The complexity is $O(l_{\text{r}}n)$. We focus on $f_{r}=f$ below.

\begin{comment}
To find $\frac{\partial g(\vec{y},\vec{f})}{\partial\vec{f}}$, we
calculate the differential first: 
\begin{eqnarray*}
 &  & g(\vec{y},\vec{f}+d\vec{f})-g(\vec{y},\vec{f})\\
 & = & \sum_{r}\vec{y}_{r}^{\dagger}F(\vec{f}+d\vec{f})SS^{\dagger}F(\vec{f}+d\vec{f})^{\dagger}\vec{y}_{r}\\
 &  & -\sum_{r}\vec{y}_{r}^{\dagger}F(\vec{f})SS^{\dagger}F(\vec{f})^{\dagger}\vec{y}_{r}\\
 & = & 2\Re\left[\sum_{r}\vec{y}_{r}^{\dagger}F(\vec{f})SS^{\dagger}\left(F(\vec{f})\left(j2\pi d\vec{f}J\right)\right)^{\dagger}\vec{y}_{r}\right]\\
 &  & +o(d\vec{f})\\
 & = & 4\pi\Im\left[\sum_{r}\vec{y}_{r}^{\dagger}F(\vec{f})SS^{\dagger}F(\vec{f})^{\dagger}J\vec{y}_{r}\right]d\vec{f}+o(d\vec{f}),
\end{eqnarray*}
 where we have used
\end{comment}
{} %

\section{Performance Bounds\label{sec:tv_Performance-Analysis}}

We derive the estimation performance bound Bayesian Cramér-Rao Lower
Bound (BCRLB) for the case of $\boldsymbol{f}_{r}=\boldsymbol{f}$
due to its simplicity. We see below that when $\sigma_{\boldsymbol{f}}^{-2}=0$,
the BCRLB becomes CRLB for mean square error conditioned on $\{\boldsymbol{f}=f\}$.
The bounds are not a function of $f$ as in \cite{Liu_2021TSP_MaximumPosterioriProbabilityMAPJointFineFrequencyOffsetChannelEstimationMIMOSystemsChannelsArbitraryCorrelation}.
\begin{comment}
The BCRLB is given in \cite{Tian_2013_DetectionEstimationModulationTheoryPartIDetectionEstimationFilteringTheory}
for parameter estimation with prior knowledge. 
\end{comment}
The proof of the following theorem implies that $\frac{\partial\ln\left(f_{\vec{\boldsymbol{y}}|\boldsymbol{f}}(\vec{y}|f)f_{\boldsymbol{f}}(f)\right)}{\partial f}$
and $\frac{\partial^{2}\ln\left(f_{\vec{\boldsymbol{y}}|\boldsymbol{f}}(\vec{y}|f)f_{\boldsymbol{f}}(f)\right)}{\partial f^{2}}$
are absolutely integrable with respect to $\vec{y}$ and $f$, satisfying
the conditions of BCRLB.
\begin{thm}
\label{thm:tv_CRLB} For any estimator satisfying 
\begin{eqnarray*}
\lim_{f\rightarrow\infty}\text{E}\left[\hat{\boldsymbol{f}}_{\delta}-f|\{\boldsymbol{f}=f\}\right]f_{\boldsymbol{f}}(f) & = & 0
\end{eqnarray*}
 and
\begin{eqnarray*}
\lim_{f\rightarrow-\infty}\text{E}\left[\hat{\boldsymbol{f}}_{\delta}-f|\{\boldsymbol{f}=f\}\right]f_{\boldsymbol{f}}(f) & = & 0,
\end{eqnarray*}
the mean square frequency estimation error for channel model (\ref{eq:tv_channel-model-fine-freq})
is lower bounded by the Bayesian CRLB:
\begin{eqnarray}
 &  & \text{E}\left[\left(\hat{\boldsymbol{f}}_{\delta}-\boldsymbol{f}\right)^{2}\right]\nonumber \\
 & \ge & \text{BCRLB}\nonumber \\
 & = & \frac{1}{-\text{E}\left[\frac{\partial^{2}\ln\left(f_{\vec{\boldsymbol{y}}|\boldsymbol{f}}(\vec{\boldsymbol{y}}|\boldsymbol{f})f_{\boldsymbol{f}}(\boldsymbol{f})\right)}{\partial\boldsymbol{f}^{2}}\right]},\label{eq:tv_1-E}\\
 & = & \frac{1}{\beta+\sigma_{\boldsymbol{f}}^{-2}}.\label{eq:tv_BCRLB}
\end{eqnarray}
Setting $\sigma_{\boldsymbol{f}}^{-2}=0$, the conditional mean square
error is lower bounded by the CRLB:
\begin{eqnarray}
 &  & \text{E}\left[\left(\hat{\boldsymbol{f}}_{\delta}-\boldsymbol{f}\right)^{2}|\{\boldsymbol{f}=f\}\right]\nonumber \\
 & \ge & \text{CRLB}\nonumber \\
 & = & \frac{1}{-\text{E}_{\vec{\boldsymbol{y}}|\{\boldsymbol{f}=f\}}\left[\frac{\partial^{2}\ln\left(f_{\vec{\boldsymbol{y}}|\boldsymbol{f}}(\vec{\boldsymbol{y}}|f)\right)}{\partial f^{2}}\right]},\label{eq:tv_1-E-1}\\
 & = & \frac{1}{\beta},\nonumber 
\end{eqnarray}
where 
\begin{eqnarray}
\beta & = & 8\pi^{2}\Re\left[\sum_{k=1}^{n-1}k^{2}\times\right.\nonumber \\
 &  & \left(\sum_{r,t}\sum_{t'}s_{t,k+1}s_{t',k+1}^{*}\mu_{\boldsymbol{h}_{r,t',k+1}}^{*}b_{r,t,k+1}+\right.\nonumber \\
 &  & \sum_{k_{1}=k+1}^{n}\sum_{r_{1},t_{1},r_{2},t_{2}}a_{r_{1},t_{1},k_{1},r_{2},t_{2},k_{1}-k}\times\nonumber \\
 &  & s_{t_{1},k_{1}}s_{t_{2},k_{1}-k}^{*}\sum_{t_{2}'}s_{t_{2}',k_{1}-k}\sum_{t_{1}'}s_{t_{1}',k_{1}}^{*}\times\nonumber \\
 &  & \left.\left.\left(c_{\boldsymbol{h}_{r_{1},t_{1}',k_{1}},\boldsymbol{h}_{r_{2},t_{2}',k_{1}-k}}^{*}+\mu_{\boldsymbol{h}_{r_{2},t_{2}',k_{1}-k}}\mu_{\boldsymbol{h}_{r_{1},t_{1}',k_{1}}}^{*}\right)\right)\right];\label{eq:tv_-E-general}
\end{eqnarray}
\begin{comment}
\begin{eqnarray}
 &  & \text{E}_{\vec{\boldsymbol{y}}|\boldsymbol{f}}\left[\boldsymbol{z}_{k}\right]\nonumber \\
 & = & e^{-j2\pi\boldsymbol{f}k}\sum_{r,t}\sum_{t'}s_{t,k+1}s_{t',k+1}^{*}\mu_{\boldsymbol{h}_{r,t',k+1}}^{*}b_{r,t,k+1}+\nonumber \\
 &  & e^{-j2\pi\boldsymbol{f}k}\sum_{k_{1}=k+1}^{n}\sum_{r_{1},t_{1},r_{2},t_{2}}a_{r_{1},t_{1},k_{1},r_{2},t_{2},k_{1}-k}\times\nonumber \\
 &  & s_{t_{1},k_{1}}s_{t_{2},k_{1}-k}^{*}\sum_{t_{2}'}s_{t_{2}',k_{1}-k}\sum_{t_{1}'}s_{t_{1}',k_{1}}^{*}\times\nonumber \\
 &  & \left(c_{\boldsymbol{h}_{r_{1},t_{1}',k_{1}},\boldsymbol{h}_{r_{2},t_{2}',k_{1}-k}}^{*}+\mu_{\boldsymbol{h}_{r_{2},t_{2}',k_{1}-k}}\mu_{\boldsymbol{h}_{r_{1},t_{1}',k_{1}}}^{*}\right),\ k\ne0.\label{eq:tv_Ertheta-1}
\end{eqnarray}
\end{comment}
$a_{r_{1},t_{1},k_{1},r_{2},t_{2},k_{1}-k}$ is given in \prettyref{eq:tv_def-A};
$b_{r,t,k+1}$ is given in \prettyref{eq:tv_def-b}, $c_{\boldsymbol{h}_{r_{1},t_{1}',k_{1}},\boldsymbol{h}_{r_{2},t_{2}',k_{1}-k}}^{*}$
is defined in \prettyref{eq:tv_cov-h}.
\end{thm}
The proof is given in Appendix \ref{sec:tv_Proof-CRLB}.
\begin{rem}
For the special case of zero mean, spatially uncorrelated, wide-sense
stationary fading, the BCRLB can be simplified and calculated as a
closed form function of the power spectral density of the fading because
of $A$'s calculation as in Remark \prettyref{rem:tv_A-special-case}. 
\end{rem}
 
\begin{rem}
Pilot/Training Signal Design for CFO and Channel Estimation: \emph{Orthogonality:}
The BCRLB can guide the design of the pilot signals for frequency
estimation. The pilot signal is also used for channel estimation.
Since in general it is not practical to design pilot signals for each
specific channel correlation, one should design it for i.i.d. channel
coefficients. It is easy to prove that the optimal pilot for channel
estimation for i.i.d. channel satisfies $S^{\dagger}S=\frac{n\rho}{l_{\text{t}}}I$,
as long as $n\ge l_{\text{t}}$ so that the pilot signals are orthogonal
across transmit antennas. This is the same as in the time invariant
case.\emph{ Time Spread:} We gain insight from a special case. If
the fading is correlated across $l_{\text{t}}$ symbol time but uncorrelated
beyond that. Then, we can perform frequency estimation with the TD
pilot but not with the periodic pilot because the CFO related phase
can not be distinguished from the phase of the fading. When the fading
correlation over time increases, the advantage of spreading the pilot
symbol over time like the periodic pilot will take effect as in the
time invariant fading case of \cite{Liu_2021TSP_MaximumPosterioriProbabilityMAPJointFineFrequencyOffsetChannelEstimationMIMOSystemsChannelsArbitraryCorrelation}. 
\end{rem}

%% file: synchronization_fine_freq_time_varying_Results_v4.tex
\section{Simulation Results\label{sec:tv_Simulation-Results}}

We use a simple time varying fading channel model to examine the impact
of time correlation coefficient on the performance. Some results are
not obvious.

\begin{comment}
We consider a general $2\times2$ MIMO system with as the scenario.
For simplicity, BPSK modulation is employed for simulations. To acquire
accurate simulation results, we conduct Monte-Carlo simulation and
gather the mean square error (MSE) over the simulations. With these
given parameters, CRLBs with different SNRs are calculated to compare
with simulation results as the performance guideline. For convenience,
we normalized the CFO, where the normalized CFO is set to be a proportion
of the samplilng rate. Simulations for both periodic pilot and TD
pilot are conducted and compared under same conditions. As can be
seen clearly from the figures, the MSE of proposed fine frequency
synchronization method runs close to the derived CRLB. In addition,
the TD pilot has larger range for CFO acquisition and it also has
the potential of handling multiple CFOs, while the MSE of TD pilot
is larger than that of periodic pilot in an MIMO system.
\end{comment}

\subsection{Time Varying Fading Model}

There are two popular methods to produce time varying fading channels
for simulation, ray tracing \cite{Chen:00,_Specification36211} and
ARMA filtering of a Gaussian process \cite{Blostein_2007G2-GTC_ARMASynthesisFadingChannelsApplicationGenerationDynamicMIMOChannels,Blostein_2008TWC_ARMASynthesisFadingChannels,Brossier_2012SP_UseFirstorderAutoregressiveModelingRayleighFlatFadingChannelEstimationKalmanFilter}.
\begin{comment}
Ray tracing is easy and realistic, using the Jake's model \cite{Jakes:74}
for SISO fading channels, the Chen, \emph{et al}., model for space-time
fading \cite{Chen:00}, and the 3GPP standard specified method \cite{_SpatialChannelModelMIMOSimulationsRayBasedSimulatorBased3GPPTR25996V610,_Specification36211}. 
\end{comment}
{} We use a simple ARMA model so that the time correlation coefficient
can be specified by a single parameter $\rho_{h}$ to examine its
impact on BCRLB.

The ARMA model can be produced using
\begin{eqnarray*}
\sum_{n=0}^{\bar{n}}a_{n}\boldsymbol{h}_{r,t,k-n} & = & \sum_{m=0}^{\bar{n}}b_{n}\boldsymbol{w}_{r,t,k-m}^{h}+c\mu_{\boldsymbol{h}_{r,t}},
\end{eqnarray*}
 a rational filtering of the Gaussian process $\boldsymbol{w}_{r,t,k-m}^{h}$.
This can be further generalized by expanding the summations over $r,t$. 

We use a simple AR(1) model 
\begin{eqnarray*}
\boldsymbol{h}_{r,t,k} & = & \rho_{h}\boldsymbol{h}_{r,t,k-1}+\sqrt{1-\rho_{h}^{2}}\boldsymbol{w}_{r,t,k}^{h}+(1-\rho_{h})\mu_{\boldsymbol{h}_{r,t}},
\end{eqnarray*}
 where $\left[\left[\boldsymbol{w}_{r,t,k}^{h}\right]_{r}\right]_{t}\sim\mathcal{CN}\left(\vec{0},\left[\left[c_{r,t,r',t'}\right]_{r,r'}\right]_{t,t'}\right)$
is i.i.d. over $k$; and
\begin{eqnarray*}
\boldsymbol{h}_{r,t,1} & = & \boldsymbol{w}_{r,t,1}^{h}+\mu_{\boldsymbol{h}_{r,t}}.
\end{eqnarray*}
 It is easy to calculate that $\mathbb{E}\left[\boldsymbol{h}_{r,t,k}\right]=\mu_{\boldsymbol{h}_{r,t}}$,
$\forall k$. 

We analyze the covariance as follows. If $k>k'$, then
\begin{eqnarray*}
 &  & \text{Cov}\left[\boldsymbol{h}_{r,t,k},\boldsymbol{h}_{r',t',k'}\right]\\
 & = & \mathbb{E}\left[\left(\boldsymbol{h}_{r,t,k}-\mu_{\boldsymbol{h}_{r,t}}\right)\left(\boldsymbol{h}_{r',t',k'}-\mu_{\boldsymbol{h}_{r',t'}}\right)^{*}\right]\\
 & = & \mathbb{E}[\left(\rho_{h}\left(\boldsymbol{h}_{r,t,k-1}-\mu_{\boldsymbol{h}_{r,t}}\right)+\sqrt{1-\rho_{h}^{2}}\boldsymbol{w}_{r,t,k}^{h}\right)\\
 &  & \left(\boldsymbol{h}_{r',t',k'}-\mu_{\boldsymbol{h}_{r',t'}}\right)^{*}]\\
 & = & \rho_{h}\text{Cov}\left[\boldsymbol{h}_{r,t,k-1},\boldsymbol{h}_{r',t',k'}\right].
\end{eqnarray*}
If $k<k'$, then
\begin{eqnarray*}
 &  & \text{Cov}\left[\boldsymbol{h}_{r,t,k},\boldsymbol{h}_{r',t',k'}\right]\\
 & = & \text{Cov}\left[\boldsymbol{h}_{r',t',k'},\boldsymbol{h}_{r,t,k}\right]^{*}\\
 & = & \rho_{h}\text{Cov}\left[\boldsymbol{h}_{r',t',k'-1},\boldsymbol{h}_{r,t,k}\right]^{*}\\
 & = & \rho_{h}\text{Cov}\left[\boldsymbol{h}_{r,t,k},\boldsymbol{h}_{r',t',k'-1}\right].
\end{eqnarray*}
If $k=k'$=2,3,..., then 
\begin{eqnarray*}
 &  & \text{Cov}\left[\boldsymbol{h}_{r,t,k},\boldsymbol{h}_{r',t',k}\right]\\
 & = & \mathbb{E}\left[\left(\boldsymbol{h}_{r,t,k}-\mu_{\boldsymbol{h}_{r,t}}\right)\left(\boldsymbol{h}_{r',t',k}-\mu_{\boldsymbol{h}_{r',t'}}\right)^{*}\right]\\
 & = & \mathbb{E}[\left(\rho_{h}\left(\boldsymbol{h}_{r,t,k-1}-\mu_{\boldsymbol{h}_{r,t}}\right)+\sqrt{1-\rho_{h}^{2}}\boldsymbol{w}_{r,t,k}^{h}\right)\\
 &  & \left(\rho_{h}\left(\boldsymbol{h}_{r',t',k-1}-\mu_{\boldsymbol{h}_{r',t'}}\right)+\sqrt{1-\rho_{h}^{2}}\boldsymbol{w}_{r',t',k}^{h}\right)^{*}]\\
 & = & \rho_{h}^{2}\text{Cov}\left[\boldsymbol{h}_{r,t,k-1},\boldsymbol{h}_{r',t',k-1}\right]+\left(1-\rho_{h}^{2}\right)\text{Cov}\left[\boldsymbol{w}_{r,t,k}^{h},\boldsymbol{w}_{r',t',k}^{h}\right]\\
 & = & \rho_{h}^{2}c_{r,t,r',t'}+\left(1-\rho_{h}^{2}\right)c_{r,t,r',t'}\\
 & = & c_{r,t,r',t'}.
\end{eqnarray*}
Therefore, 
\begin{eqnarray*}
\text{Cov}\left[\boldsymbol{h}_{r,t,k},\boldsymbol{h}_{r',t',k'}\right] & = & c_{\boldsymbol{h}_{r,t,k},\boldsymbol{h}_{r',t',k'}}\\
 & = & \rho_{h}^{\left|k-k'\right|}c_{r,t,r',t'}.
\end{eqnarray*}
Consequently, the model's spatial and time correlation coefficient
is separable. The mean and spatial correlation are constant.

\subsection{Results}

\emph{Simulation Parameters:} Unless noted otherwise, the default
simulation parameters are as follows. Number of antennas: $l_{t}=4$,
$l_{r}=4$; Pilot length: $n=20$ symbols; CFO distribution: $\boldsymbol{f}_{\delta}$$\sim\mathcal{N}(\mu_{\boldsymbol{f}_{\delta}},\sigma_{\boldsymbol{f}_{\delta}}^{2})$
where $\mu_{\boldsymbol{f}_{\delta}}=0.1$, $\sigma_{\boldsymbol{f}_{\delta}}^{2}=10^{-5}$;
Spatially uncorrelated and zero mean fading.

\begin{figure}
\begin{centering}
\includegraphics[width=0.8\columnwidth]{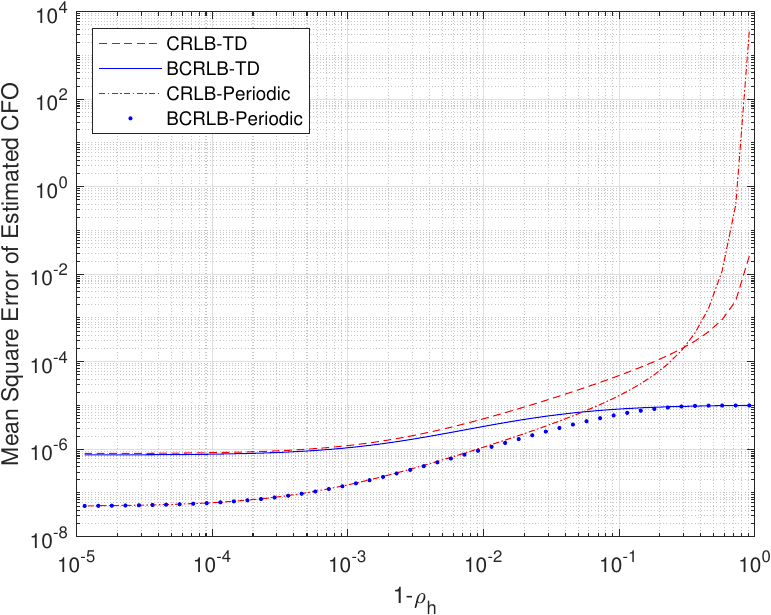}
\par\end{centering}
\caption{\label{fig:BCRLB_rho_h_SNR20} Comparison of the TD and Periodic pilots'
CRLB and BCRLB as a function of time correlation coefficient: SNR=20
dB.}
\end{figure}

\begin{figure}
\begin{centering}
\includegraphics[width=0.8\columnwidth]{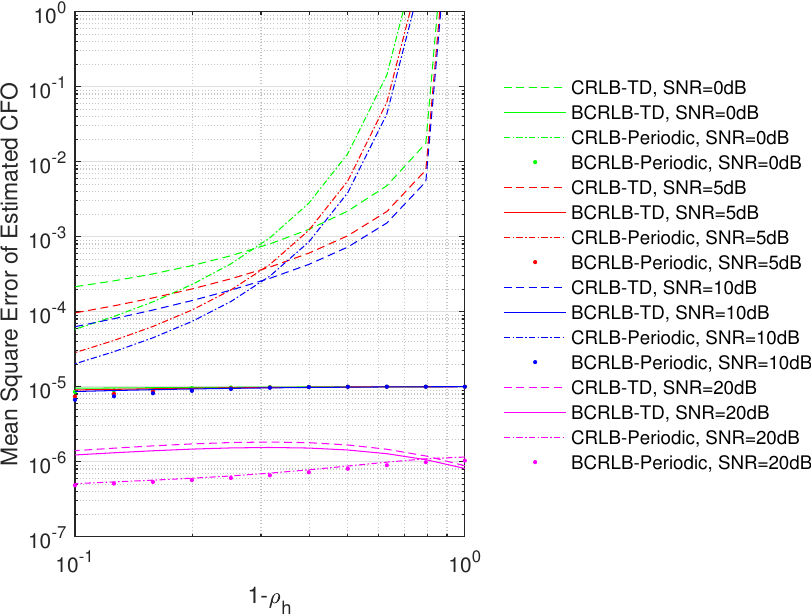}
\par\end{centering}
\caption{\label{fig:BCRLB_rho_h_diff_SNR_test_crossing}Crossing points of
the performance of Periodic and TD pilots at different SNR: Nonzero
mean and spatially correlated fading.}
\end{figure}

\emph{Good pilot signals depend on time correlation coefficient:}
Figure \ref{fig:BCRLB_rho_h_SNR20} shows that the CRLB of TD pilot
is better than that of Periodic pilot for time correlation coefficient
less than $\rho_{\boldsymbol{h}}=0.7$. It is due to the TD pilot's
sampling of consecutive symbol time, where the fading has higher time
correlation coefficient than that of every $l_{t}$ symbols of the
Periodic pilot case. With prior knowledge, the difference between
BCRLBs of TD and periodic pilots is not noticeable at low correlation
region. The crossing points of the performance of the Periodic and
TD pilots happen at less time correlation coefficient when the SNR
increases as shown in Figure \ref{fig:BCRLB_rho_h_diff_SNR_test_crossing}.
Therefore, the pilot signal design need to consider typical time correlation
coefficient in applications with ML CFO estimation. 

\begin{figure}
\begin{centering}
\includegraphics[width=0.8\columnwidth]{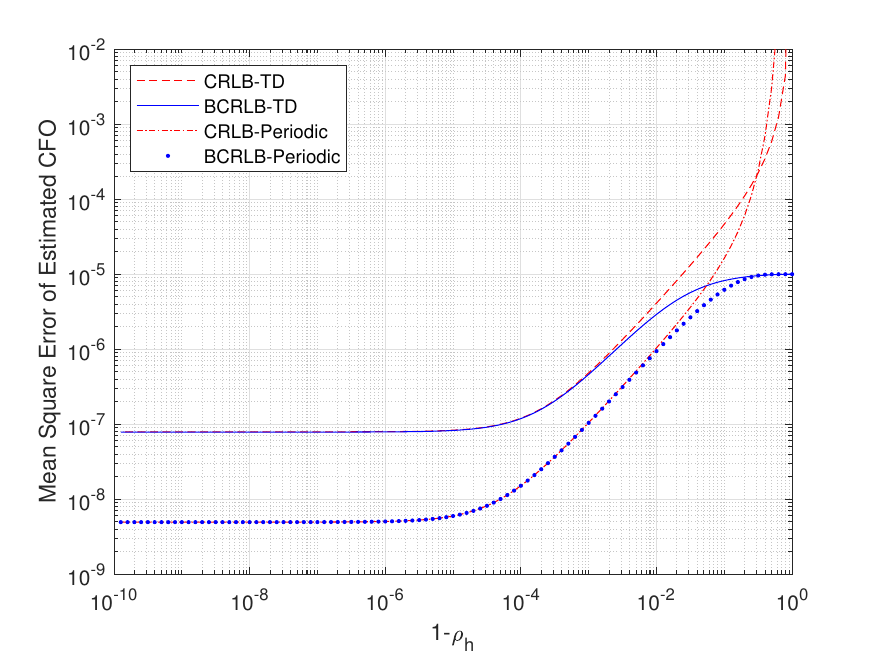}
\par\end{centering}
\caption{\label{fig:time_correlation_close_to_1} Examination of how CRLB/BCRLB
deteriorates as time correlation coefficient decreases by a small
amount: SNR=30 dB.}
\end{figure}

\emph{Performance deteriorates even with a very small loss of time
correlation coefficient:} Figure \ref{fig:time_correlation_close_to_1}
shows that at high SNR, the CRLB/BCRLB starts to be notably worse
at $\rho_{\boldsymbol{h}}=1-10^{-4}$ than at $\rho_{\boldsymbol{h}}=1$.
Therefore, a loss of time correlation in fading is a significant design
consideration for CFO estimation performance. 

\begin{figure}
\begin{centering}
\includegraphics[width=0.8\columnwidth]{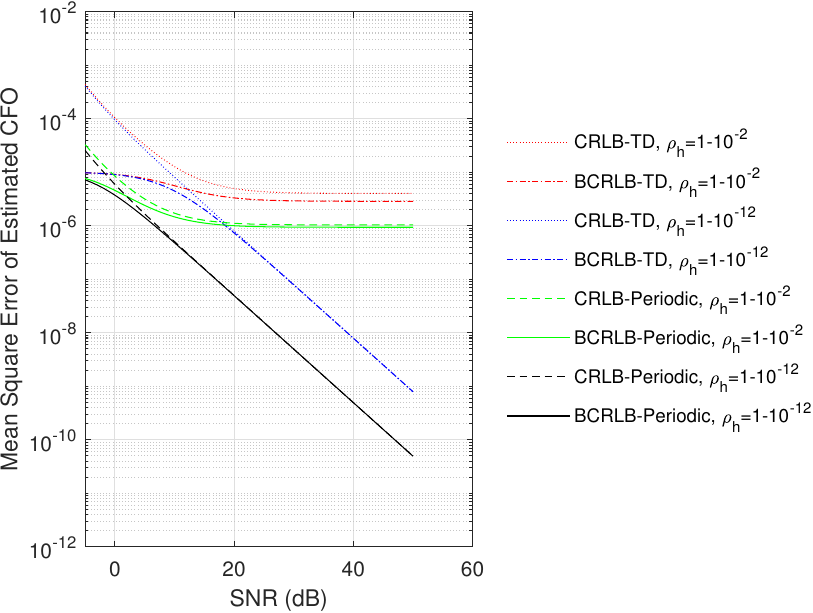}
\par\end{centering}
\caption{\label{fig:error_floor_time_correlation-7}CRLB/BCRLB error floors
even for a $\rho_{\boldsymbol{h}}=0.99$ correlation.}
\end{figure}

\emph{The loss of time correlation coefficient causes error floor:}
Figure \ref{fig:error_floor_time_correlation-7} shows that when the
time correlation coefficient of the fading decreases from 1 to 0.99,
the performance bounds start to deteriorate at low SNR and meet an
error floor starting at moderate SNR. Thus, the time correlation coefficient
can have a significant impact on the performance. 

\begin{figure}
\begin{centering}
\includegraphics[width=0.8\columnwidth]{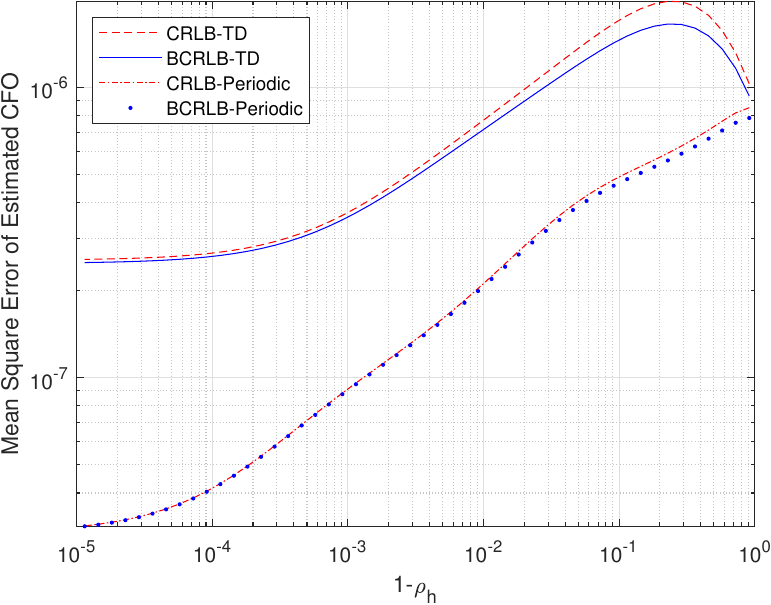}
\par\end{centering}
\caption{\label{fig:BCRLB_rho_h_nonmonotone_TD}Non-monotonicity of CRLB/BCRLB
of TD pilot as a function of time correlation coefficient: Nonzero
mean and spatially correlated fading, SNR=20 dB.}
\end{figure}

\begin{figure}
\begin{centering}
\includegraphics[width=0.8\columnwidth]{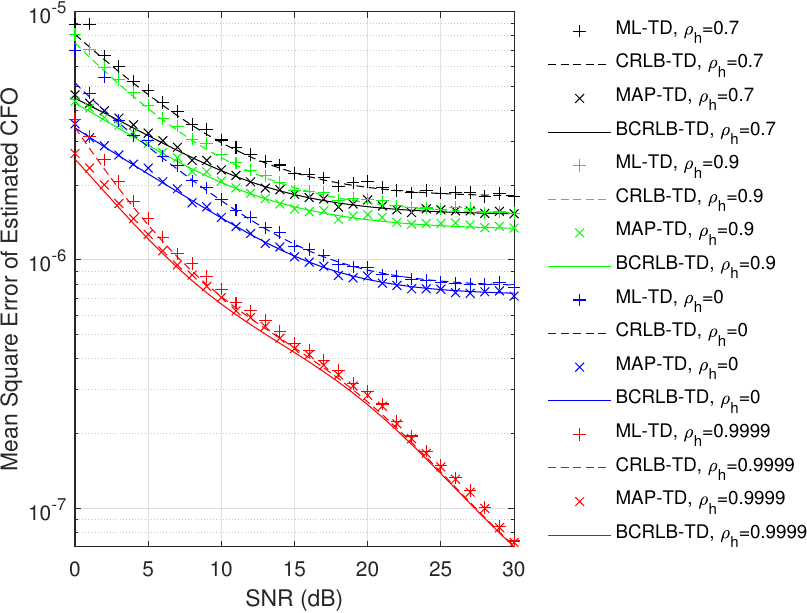}
\par\end{centering}
\caption{\label{fig:BCRLB_SNR_diff_corr_test_monotone_TD} MSE and CRLB/BCRLB
as a function of SNR for different time correlation coefficients for
TD Pilot.}
\end{figure}

\begin{figure}
\begin{centering}
\includegraphics[width=0.8\columnwidth]{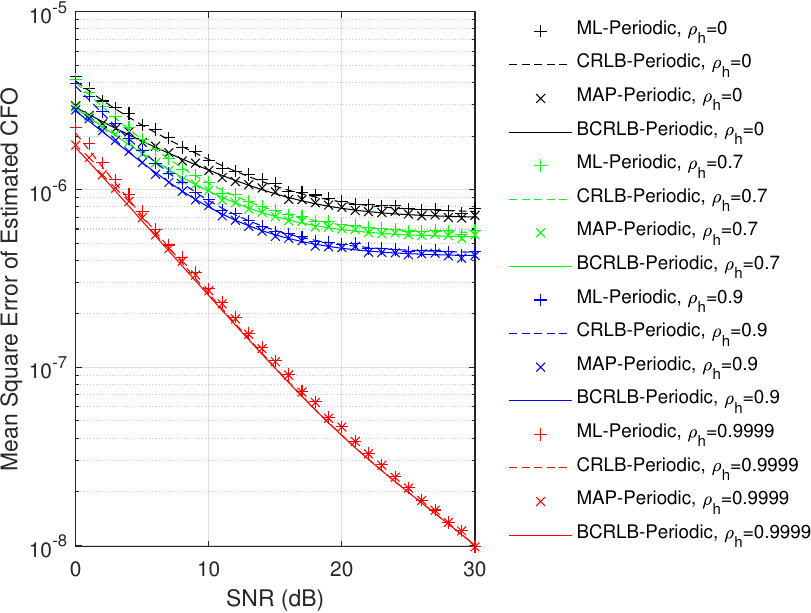}
\par\end{centering}
\caption{\label{fig:BCRLB_SNR_diff_corr_test_monotone_periodic}MSE and CRLB/BCRLB
as a function of SNR for different time correlation coefficients for
Periodic Pilot.}
\end{figure}

\begin{figure}
\begin{centering}
\includegraphics[width=0.8\columnwidth]{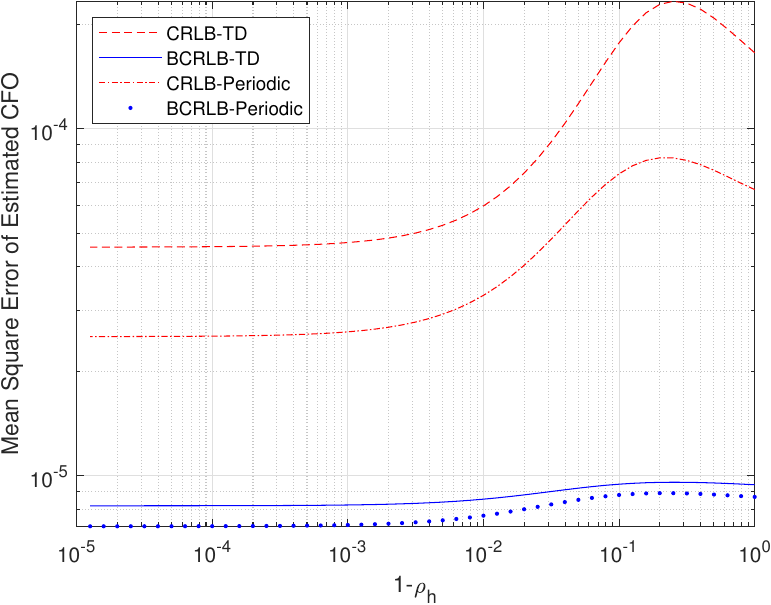}
\par\end{centering}
\caption{\label{fig:BCRLB_rho_h_nonmonotone_both}Non-monotonicity of CRLB/BCRLB
of TD and Periodic pilot as a function of time correlation coefficient:
Nonzero mean and spatially correlated fading, SNR=-5 dB, $l_{t}=2$,
$l_{r}=2$.}
\end{figure}

\emph{Time diversity of the fading may benefit the performance:} It
is intuitive to think that the CFO estimation performance deteriorates
as the time correlation coefficient decreases. But Figure \ref{fig:BCRLB_rho_h_nonmonotone_TD}
shows unexpectedly that the CRLB/BCRLB may \emph{not} be a \emph{monotone}
function of the time correlation coefficient and may improve when
the time correlation coefficient decreases to zero if the fading has
nonzero mean, due to the benefit of time diversity. The correctness
of the non-monotone bounds is verified by that the simulation results
achieve the CRLB and BCRLB for various correlations in Figure \ref{fig:BCRLB_SNR_diff_corr_test_monotone_TD}
and \ref{fig:BCRLB_SNR_diff_corr_test_monotone_periodic}. They also
demonstrate the near optimal performance of the universal algorithm.
The CRLBs/BCRLBs of both the TD pilot and Periodic pilot may not be
a monotone function of time correlation coefficient as shown in Figure
\ref{fig:BCRLB_rho_h_nonmonotone_both}. Thus, CFO estimation in time
uncorrelated fading is not hopeless when the fading has nonzero mean.
On the other hand, for zero mean and time uncorrelated fading channels,
the phase changes due to CFO is not distinguishable from the phase
changes due to fading. As a result, the CFO estimation is not possible.

%% file: synchronization_fine_freq_time_varying_conclusion_v4.tex
\section{Conclusion\label{sec:tv_Conclusion}}

In this work, the solution of the joint MAP estimation of channel
states and the frequency offset in time varying and spatially correlated
fading channels is provided. When $\boldsymbol{f}_{r,t}=\boldsymbol{f}_{r}$,
the solution is separable with an individual MAP estimation of the
CFOs with channel statistic information first. A near closed form
and near optimal algorithm is given for the case of $\boldsymbol{f}_{r}=\boldsymbol{f}$.
The Bayesian Cramér-Rao Lower bounds (BCRLB) are derived in closed
form for the frequency offset estimation with prior knowledge for
the case of $\boldsymbol{f}_{r}=\boldsymbol{f}$. Numeric results
are provided to show that a small decrease of time correlation coefficient
causes significant performance deterioration but the time diversity
may benefit the performance if the fading has nonzero means.

%% file: synchronization_fine_freq_time_varying_apdx_v6.tex
\section{\label{sec:tv_Proof-of-MAP-channel}Proof of Theorem \ref{thm:MAP-channel}
of the MAP Estimator of the Channel}

\begin{comment}
--------proof
\end{comment}

The channel model implies $\vec{\boldsymbol{h}},\vec{\boldsymbol{y}}$
are jointly Gaussian conditioned on $\vec{\boldsymbol{f}}$. Therefore,
\begin{eqnarray*}
f_{\vec{\boldsymbol{h}}|\vec{\boldsymbol{y}},\vec{\boldsymbol{f}}}(\vec{h}|\vec{y},\vec{f}) & = & \mathcal{CN}\left(\hat{\vec{h}}_{\text{MMSE}}(\vec{y},\vec{f}),\Sigma_{\hat{\vec{\boldsymbol{h}}}_{\text{MMSE}}}\right)(\vec{h}),
\end{eqnarray*}
 where $\mathcal{CN}\left(\vec{\mu},\Sigma\right)(\vec{x})=\frac{1}{\det\left(\pi\Sigma\right)}e^{-\left(x-\vec{\mu}\right)^{\dagger}\Sigma^{-1}\left(x-\vec{\mu}\right)}$
denotes the circularly symmetric complex Gaussian density function;
$\hat{\vec{h}}_{\text{MMSE}}(\vec{y},\vec{f})$ is the MMSE estimate
of $\vec{\boldsymbol{h}}$ and $\Sigma_{\hat{\vec{\boldsymbol{h}}}_{\text{MMSE}}}$
is the MMSE estimation error covariance, which does not depend on
$\vec{y}$ or $\vec{f}$, as shown below.

To calculate $\hat{\vec{h}}_{\text{MMSE}}(\vec{y},\vec{f})$, find
the mean of $\vec{\boldsymbol{y}}$ given the frequency offset as
\begin{eqnarray}
\vec{\mu}_{\vec{\boldsymbol{y}}|\vec{\boldsymbol{f}}} & = & \text{E}\left[\vec{\boldsymbol{y}}|\{\vec{\boldsymbol{f}}=\vec{f}\}\right]\nonumber \\
 & = & \text{E}\left[\grave{X}\vec{\boldsymbol{h}}+\vec{\boldsymbol{n}}\right]\nonumber \\
 & = & \grave{X}\vec{\mu}_{\vec{\boldsymbol{h}}}.\label{eq:tv_Ey-f}
\end{eqnarray}
By the MMSE estimation theory, the the MMSE estimate is %
\begin{comment}
\begin{eqnarray*}
\hat{\vec{h}}_{\text{MMSE}}(\vec{y},\vec{f}) & = & (\mathbf{H}^{\dagger}\boldsymbol{\Omega}^{-1}\mathbf{H}+\mathbf{P}^{-1})^{-1}\mathbf{H}^{\dagger}\boldsymbol{\Omega}^{-1}\mathbf{y}
\end{eqnarray*}
\end{comment}
\begin{eqnarray}
\hat{\vec{h}}_{\text{MMSE}}(\vec{y},\vec{f}) & = & \underset{A}{\underbrace{\left(\grave{X}^{\dagger}\grave{X}+\Sigma_{\vec{\boldsymbol{h}}}^{-1}\right)^{-1}}}\grave{X}^{\dagger}\left(\vec{y}-\grave{X}\vec{\mu}_{\vec{\boldsymbol{h}}}\right)+\vec{\mu}_{\vec{\boldsymbol{h}}}\nonumber \\
 & = & A\grave{X}^{\dagger}\left(\vec{y}-\grave{X}\vec{\mu}_{\vec{\boldsymbol{h}}}\right)+\vec{\mu}_{\vec{\boldsymbol{h}}}\nonumber \\
 & = & A\grave{X}^{\dagger}\vec{y}+\vec{b},\label{eq:h_MMSE}
\end{eqnarray}
where
\begin{eqnarray*}
\grave{X}^{\dagger}\grave{X} & = & \left[\left[\left[e^{-j2\pi\boldsymbol{f}_{r,t_{1}}(k-1)}s_{t_{1},k}^{*}\right]_{t_{1}}\left[e^{j2\pi\boldsymbol{f}_{r,t_{2}}(k-1)}s_{t_{2},k}\right]_{1,t_{2}}\right]_{k,k}\right]_{r,r}\\
 & = & \left[\left[\left[e^{j2\pi\left(\boldsymbol{f}_{r,t_{2}}-\boldsymbol{f}_{r,t_{1}}\right)(k-1)}s_{t_{1},k}^{*}s_{t_{2},k}\right]_{t_{1},t_{2}}\right]_{k,k}\right]_{r,r},
\end{eqnarray*}
\begin{eqnarray*}
\grave{S} & = & \left[\left[\left[s_{t,k}\right]_{1,t}\right]_{k,k}\right]_{r,r},
\end{eqnarray*}
\begin{eqnarray*}
\grave{S}^{\dagger}\grave{S} & = & \left[\left[\left[s_{t',k}^{*}\right]_{t'}\right]_{k,k}\right]_{r,r}\left[\left[\left[s_{t,k}\right]_{1,t}\right]_{k,k}\right]_{r,r}\\
 & = & \left[\left[\left[s_{t',k}^{*}s_{t,k}\right]_{t',t}\right]_{k,k}\right]_{r,r},
\end{eqnarray*}
{\tiny{}
\begin{eqnarray}
 &  & A\nonumber \\
 & = & \left(\grave{X}^{\dagger}\grave{X}+\Sigma_{\vec{\boldsymbol{h}}}^{-1}\right)^{-1}\\
 & = & \left[\left[\left[a_{r_{1},t_{1},k_{1},r_{2},t_{2},k_{2}}\right]_{t_{1},t_{2}}\right]_{k_{1},k_{2}}\right]_{r_{1},r_{2}}\nonumber \\
 & \triangleq & \begin{cases}
\left(\grave{S}^{\dagger}\grave{S}+\Sigma_{\vec{\boldsymbol{h}}}^{-1}\right)^{-1} & \boldsymbol{f}_{r,t}=\boldsymbol{f}_{r},\forall t\\
\left(\left[\left[\left[e^{j2\pi\left(\boldsymbol{f}_{r,t_{2}}-\boldsymbol{f}_{r,t_{1}}\right)(k-1)}s_{t_{1},k}^{*}s_{t_{2},k}\right]_{t_{1},t_{2}}\right]_{k,k}\right]_{r,r}+\Sigma_{\vec{\boldsymbol{h}}}^{-1}\right)^{-1} & \text{else}
\end{cases},\label{eq:tv_def-A}
\end{eqnarray}
}
\begin{eqnarray}
\vec{b} & = & \left[\left[\left[b_{r,t,k}\right]_{t}\right]_{k}\right]_{r}\nonumber \\
 & \triangleq & \left(I-A\grave{X}^{\dagger}\grave{X}\right)\vec{\mu}_{\vec{\boldsymbol{h}}}\nonumber \\
 & = & \left(I+\Sigma_{\vec{\boldsymbol{h}}}\grave{X}^{\dagger}\grave{X}\right)^{-1}\vec{\mu}_{\vec{\boldsymbol{h}}}\\
 & = & \begin{cases}
\left(I-A\grave{S}^{\dagger}\grave{S}\right)\vec{\mu}_{\vec{\boldsymbol{h}}} & \boldsymbol{f}_{r,t}=\boldsymbol{f}_{r},\forall t\\
\left(I+\Sigma_{\vec{\boldsymbol{h}}}\grave{X}^{\dagger}\grave{X}\right)\vec{\mu}_{\vec{\boldsymbol{h}}} & \text{else}
\end{cases},\label{eq:tv_def-b}
\end{eqnarray}
\begin{comment}
From MMSE estimation theory, the MMSE estimate of the channel is
\begin{eqnarray*}
\hat{\vec{h}}_{\text{MMSE}}(\vec{y},\vec{f}) & = & \left[\left[X^{\dagger}X+\frac{1}{\sigma_{\boldsymbol{h}}^{2}}I_{l_{\text{t}}}\right]^{-1}X^{\dagger}\vec{y}_{r}\right]_{r}\\
 & = & \left[\left[S^{\dagger}S+\frac{1}{\sigma_{\boldsymbol{h}}^{2}}I_{l_{\text{t}}\times l_{\text{t}}}\right]^{-1}S^{\dagger}F^{\dagger}\vec{y}_{r}\right]_{r}\\
 & = & \left[\left[\frac{n\rho}{l_{\text{t}}}I_{l_{\text{t}}\times l_{\text{t}}}+\frac{1}{\sigma_{\boldsymbol{h}}^{2}}I_{l_{\text{t}}\times l_{\text{t}}}\right]^{-1}S^{\dagger}F^{\dagger}\vec{y}_{r}\right]_{r}\\
 & = & \left(\frac{n\rho}{l_{\text{t}}}+\frac{1}{\sigma_{\boldsymbol{h}}^{2}}\right)^{-1}\left[S^{\dagger}F^{\dagger}\vec{y}_{r}\right]_{r}=\left[\hat{\vec{h}}_{r}\right]_{r}
\end{eqnarray*}
\end{comment}
{} and the estimation error covariance matrix is %
\begin{comment}
\begin{eqnarray*}
\Sigma_{\hat{\vec{h}}_{\text{MMSE}}} & = & \left(\grave{S}^{\dagger}\grave{S}+\Sigma_{\vec{\boldsymbol{h}}}^{-1}\right)^{-1}\\
 & = & \left(\frac{n\rho}{l_{\text{t}}}I+\Sigma_{\vec{\boldsymbol{h}}}^{-1}\right)^{-1}
\end{eqnarray*}
\end{comment}
{} %
\begin{comment}
\begin{eqnarray*}
A & = & I\\
U & = & I\\
D & = & I\\
V & = & \grave{S}^{\dagger}\grave{S}\\
C & = & \Sigma_{\vec{\boldsymbol{h}}}
\end{eqnarray*}
\begin{eqnarray*}
 &  & I-A\grave{S}^{\dagger}\grave{S}\\
 & = & \left(I+\Sigma_{\vec{\boldsymbol{h}}}\grave{S}^{\dagger}\grave{S}\right)^{-1}
\end{eqnarray*}
\end{comment}
\begin{comment}
\begin{eqnarray*}
(\mathbf{A}+\mathbf{U}\mathbf{C}\mathbf{V})^{-1} & = & \mathbf{A}^{-1}-\mathbf{A}^{-1}\mathbf{U}(\mathbf{C}^{-1}+\mathbf{V}\mathbf{A}^{-1}\mathbf{U})^{-1}\mathbf{V}\mathbf{A}^{-1},
\end{eqnarray*}
\begin{eqnarray*}
\mathbf{BC(ABC+D)}^{-1} & = & (\mathbf{C}\mathbf{D}^{-1}\mathbf{A}+\mathbf{B}^{-1})^{-1}\mathbf{C}\mathbf{D}^{-1},
\end{eqnarray*}
\end{comment}
\begin{eqnarray}
\Sigma_{\hat{\vec{\boldsymbol{h}}}_{\text{MMSE}}} & = & A,\label{eq:Sigma_hMMSE}
\end{eqnarray}
{} which is not a function of $\vec{f}$ when the frequencies only deffer
by the receive antennas, i.e., $\boldsymbol{f}_{r,t_{2}}=\boldsymbol{f}_{r,t_{1}},\ \forall t_{1},t_{2}$.
\begin{comment}
\begin{eqnarray*}
\Sigma_{\hat{\vec{h}}_{\text{MMSE}}} & = & \left(\frac{\rho n}{l_{\text{t}}}+\frac{1}{\sigma_{\boldsymbol{h}}^{2}}\right)^{-1}I_{l_{\text{t}}l_{\text{r}}}.
\end{eqnarray*}
\end{comment}
{} Due to the Gaussian distribution, the solution to (\ref{eq:tv_max-h})
is the MMSE estimate.

\begin{comment}
-----proof
\end{comment}

\section{\label{sec:tv_Proof-of-MAP}Proof of Theorem \ref{thm:tv_MAP-f}
of the MAP Estimator of the Frequency}

Since $\det(\pi\Sigma_{\vec{\boldsymbol{y}}|\vec{\boldsymbol{f}}})$
is not a function of $\vec{f}$, we can maximize the exponent in \prettyref{eq:tv_jointpdff}
as
\begin{eqnarray}
\hat{\vec{f}} & = & \arg\max_{\vec{f}}f_{\vec{\boldsymbol{y}}|\vec{\boldsymbol{f}}}(\vec{y}|\vec{f})f_{\vec{\boldsymbol{f}}}(\vec{f})\nonumber \\
 & = & \arg\max_{\vec{f}}-\left(\vec{y}-\vec{\mu}_{\vec{\boldsymbol{y}}|\vec{\boldsymbol{f}}}\right)^{\dagger}\Sigma_{\vec{\boldsymbol{y}}|\vec{\boldsymbol{f}}}^{-1}\left(\vec{y}-\vec{\mu}_{\vec{\boldsymbol{y}}|\vec{\boldsymbol{f}}}\right)\nonumber \\
 &  & -\frac{1}{2}\left(\vec{f}-\vec{\mu}_{\vec{\boldsymbol{f}}}\right)^{\dagger}\Sigma_{\vec{\boldsymbol{f}}}^{-1}\left(\vec{f}-\vec{\mu}_{\vec{\boldsymbol{f}}}\right)\label{eq:tv_pdf-exponent}
\end{eqnarray}
\begin{comment}
\begin{eqnarray}
\hat{\vec{f}} & = & \arg\max_{\vec{f}}f_{\vec{\boldsymbol{y}}|\vec{\boldsymbol{f}}}(\vec{y}|\vec{f})f_{\vec{\boldsymbol{f}}}(\vec{f})\nonumber \\
 & = & \arg\max_{\vec{f}}-\left(\vec{y}-\vec{\mu}_{\vec{\boldsymbol{y}}|\vec{\boldsymbol{f}}}\right)^{\dagger}\Sigma_{\vec{\boldsymbol{y}}|\vec{\boldsymbol{f}}}^{-1}\left(\vec{y}-\vec{\mu}_{\vec{\boldsymbol{y}}|\vec{\boldsymbol{f}}}\right)\nonumber \\
 &  & -\frac{1}{2}\sigma_{\vec{\boldsymbol{f}}}^{-2}\left|\vec{f}-\vec{\mu}_{\vec{\boldsymbol{f}}}\right|^{2}\label{eq:tv_pdf-exponent-1}\\
 & = & \arg\max_{\vec{f}}2\Re\left[\left\langle \grave{X}\vec{y},\vec{\mu}_{\vec{\boldsymbol{h}}}\right\rangle \right]-\nonumber \\
 &  & \frac{n\rho}{l_{\text{t}}}2\Re\left[\left\langle \grave{X}^{\dagger}\vec{y},\left(\Sigma_{\vec{\boldsymbol{h}}}^{-1}+\frac{n\rho}{l_{\text{t}}}I\right)^{-1}\vec{\mu}_{\vec{\boldsymbol{h}}}\right\rangle \right]\nonumber \\
 &  & +\left(\grave{X}^{\dagger}\vec{y}\right)^{\dagger}\left(\Sigma_{\vec{\boldsymbol{h}}}^{-1}+\frac{n\rho}{l_{\text{t}}}I\right)^{-1}\left(\grave{X}^{\dagger}\vec{y}\right)\nonumber \\
 &  & -\sigma_{\vec{\boldsymbol{f}}}^{-2}\left|\vec{f}-\vec{\mu}_{\vec{\boldsymbol{f}}}\right|^{2}\label{eq:tv_rid-terms-1-1}\\
 & = & \arg\max_{\vec{f}}g(\vec{y},\vec{f}),\nonumber 
\end{eqnarray}
\end{comment}
{} 

We calculate the terms in \prettyref{eq:tv_pdf-exponent} below. %
\begin{comment}
\begin{eqnarray*}
 &  & g(\vec{y},\vec{f})\\
 & \triangleq & 2\Re\left[\left\langle \grave{X}^{\dagger}\vec{y},\left(I-\frac{n\rho}{l_{\text{t}}}\left(\Sigma_{\vec{\boldsymbol{h}}}^{-1}+\frac{n\rho}{l_{\text{t}}}I\right)^{-1}\right)\vec{\mu}_{\vec{\boldsymbol{h}}}\right\rangle \right]\\
 &  & +\left(\grave{X}^{\dagger}\vec{y}\right)^{\dagger}\left(\Sigma_{\vec{\boldsymbol{h}}}^{-1}+\frac{n\rho}{l_{\text{t}}}I\right)^{-1}\left(\grave{X}^{\dagger}\vec{y}\right)\\
\\
 &  & \left(\sum_{r}2\Re\left[\left\langle S^{\dagger}F^{\dagger}\vec{y}_{r},\vec{\mu}_{\vec{\boldsymbol{h}}_{r}}\right\rangle \right]+\sigma_{\boldsymbol{h}}^{2}\left|S^{\dagger}F^{\dagger}\vec{y}_{r}\right|^{2}\right)\\
 &  & -\left(1+\frac{\sigma_{\boldsymbol{h}}^{2}n\rho}{l_{\text{t}}}\right)\sigma_{\vec{\boldsymbol{f}}}^{-2}\left|\vec{f}-\vec{\mu}_{\vec{\boldsymbol{f}}}\right|^{2}.
\end{eqnarray*}
\end{comment}
The conditional covariance 
\begin{eqnarray}
\Sigma_{\vec{\boldsymbol{y}}|\vec{\boldsymbol{f}}}^{-1} & = & \left(\grave{X}\Sigma_{\vec{\boldsymbol{h}}}\grave{X}^{\dagger}+I\right)^{-1}\nonumber \\
 & = & \left(I-\grave{X}\left(\Sigma_{\vec{\boldsymbol{h}}}^{-1}+\grave{X}^{\dagger}\grave{X}\right)^{-1}\grave{X}^{\dagger}\right)\label{eq:tv_by-woodbury-1}\\
 & = & \left(I-\grave{X}\left(\Sigma_{\vec{\boldsymbol{h}}}^{-1}+\grave{S}^{\dagger}\grave{S}\right)^{-1}\grave{X}^{\dagger}\right)\nonumber 
\end{eqnarray}
is converted by the Woodbury matrix identity. Then
\begin{eqnarray}
 &  & \left(\vec{y}-\vec{\mu}_{\vec{\boldsymbol{y}}|\vec{\boldsymbol{f}}}\right)^{\dagger}I\left(\vec{y}-\vec{\mu}_{\vec{\boldsymbol{y}}|\vec{\boldsymbol{f}}}\right)\nonumber \\
 & = & \vec{y}^{\dagger}\vec{y}+\vec{\mu}_{\vec{\boldsymbol{y}}|\vec{\boldsymbol{f}}}^{\dagger}\vec{\mu}_{\vec{\boldsymbol{y}}|\vec{\boldsymbol{f}}}-2\Re\left[\left\langle \vec{y},\vec{\mu}_{\vec{\boldsymbol{y}}|\vec{\boldsymbol{f}}}\right\rangle \right]\nonumber \\
 & = & \vec{y}^{\dagger}\vec{y}+\vec{\mu}_{\vec{\boldsymbol{h}}}^{\dagger}\grave{X}^{\dagger}\grave{X}\vec{\mu}_{\vec{\boldsymbol{h}}}-2\Re\left[\left\langle \vec{y},\grave{X}\vec{\mu}_{\vec{\boldsymbol{h}}}\right\rangle \right]\nonumber \\
 & = & \vec{y}^{\dagger}\vec{y}+\vec{\mu}_{\vec{\boldsymbol{h}}}^{\dagger}\grave{S}^{\dagger}\grave{S}\vec{\mu}_{\vec{\boldsymbol{h}}}-2\Re\left[\left\langle \grave{X}^{\dagger}\vec{y},\vec{\mu}_{\vec{\boldsymbol{h}}}\right\rangle \right],\label{eq:tv_yIy}
\end{eqnarray}
and
\begin{eqnarray}
 &  & \left(\vec{y}-\vec{\mu}_{\vec{\boldsymbol{y}}|\vec{\boldsymbol{f}}}\right)^{\dagger}\grave{X}\left(\Sigma_{\vec{\boldsymbol{h}}}^{-1}+\grave{S}^{\dagger}\grave{S}\right)^{-1}\grave{X}^{\dagger}\left(\vec{y}-\vec{\mu}_{\vec{\boldsymbol{y}}|\vec{\boldsymbol{f}}}\right)\nonumber \\
 & = & \left(\grave{X}^{\dagger}\vec{y}-\grave{S}^{\dagger}\grave{S}\vec{\mu}_{\vec{\boldsymbol{h}}}\right)^{\dagger}\left(\Sigma_{\vec{\boldsymbol{h}}}^{-1}+\grave{S}^{\dagger}\grave{S}\right)^{-1}\left(\grave{X}^{\dagger}\vec{y}-\grave{S}^{\dagger}\grave{S}\vec{\mu}_{\vec{\boldsymbol{h}}}\right)\nonumber \\
 & = & \left(\grave{X}^{\dagger}\vec{y}\right)^{\dagger}\left(\Sigma_{\vec{\boldsymbol{h}}}^{-1}+\grave{S}^{\dagger}\grave{S}\right)^{-1}\left(\grave{X}^{\dagger}\vec{y}\right)+\nonumber \\
 &  & \left(\grave{S}^{\dagger}\grave{S}\vec{\mu}_{\vec{\boldsymbol{h}}}\right)^{\dagger}\left(\Sigma_{\vec{\boldsymbol{h}}}^{-1}+\grave{S}^{\dagger}\grave{S}\right)^{-1}\left(\grave{S}^{\dagger}\grave{S}\vec{\mu}_{\vec{\boldsymbol{h}}}\right)\nonumber \\
 &  & -2\Re\left[\left(\grave{X}^{\dagger}\vec{y}\right)^{\dagger}\left(\Sigma_{\vec{\boldsymbol{h}}}^{-1}+\grave{S}^{\dagger}\grave{S}\right)^{-1}\left(\grave{S}^{\dagger}\grave{S}\vec{\mu}_{\vec{\boldsymbol{h}}}\right)\right].\label{eq:tv_yXy}
\end{eqnarray}
\begin{comment}
\begin{eqnarray*}
 &  & \left(\vec{y}-\vec{\mu}_{\vec{\boldsymbol{y}}|\vec{\boldsymbol{f}}}\right)^{\dagger}\grave{X}\grave{X}^{\dagger}\left(\vec{y}-\vec{\mu}_{\vec{\boldsymbol{y}}|\vec{\boldsymbol{f}}}\right)\\
 & = & \left|\grave{X}^{\dagger}\vec{y}-\grave{X}^{\dagger}\grave{X}\vec{\mu}_{\vec{\boldsymbol{h}}}\right|^{2}\\
 & = & \left|\grave{X}^{\dagger}\vec{y}-\frac{n\rho}{l_{\text{t}}}\vec{\mu}_{\vec{\boldsymbol{h}}}\right|^{2}\\
 & = & \left|\grave{X}^{\dagger}\vec{y}\right|^{2}+\left|\frac{n\rho}{l_{\text{t}}}\vec{\mu}_{\vec{\boldsymbol{h}}}\right|^{2}-\frac{n\rho}{l_{\text{t}}}2\Re\left[\left\langle \grave{X}^{\dagger}\vec{y},\vec{\mu}_{\vec{\boldsymbol{h}}}\right\rangle \right].
\end{eqnarray*}
\end{comment}
After discarding terms in \prettyref{eq:tv_yIy} and \prettyref{eq:tv_yXy}
that are not functions of $\vec{f}$, we obtain \prettyref{eq:tv_gyf}.
\begin{comment}
\begin{eqnarray*}
 &  & g(\vec{y},\vec{f})\\
 & \triangleq & 2\Re\left[\left\langle \grave{X}^{\dagger}\vec{y},\vec{b}\right\rangle \right]+\left(\grave{X}^{\dagger}\vec{y}\right)^{\dagger}A\left(\grave{X}^{\dagger}\vec{y}\right)\\
 &  & -\frac{1}{2}\vec{f}^{\dagger}\Sigma_{\vec{\boldsymbol{f}}}^{-1}\vec{f}+\Re\left[\vec{\mu}_{\vec{\boldsymbol{f}}}^{\dagger}\Sigma_{\vec{\boldsymbol{f}}}^{-1}\vec{f}\right]
\end{eqnarray*}
\end{comment}

To write the $g(\vec{y},\vec{f})$ in summation form, we examine the
following. The first term of it can be calculated as 
\begin{eqnarray}
 &  & \left\langle \grave{X}^{\dagger}\vec{y},\vec{b}\right\rangle \nonumber \\
 & = & \sum_{k}\sum_{r}e^{-j2\pi f_{r}(k-1)}\sum_{t}s_{t,k}^{*}y_{r,k}b_{r,t,k}^{*}.\label{eq:tv_xyb-fr-sum}
\end{eqnarray}
If $f_{r}=f$, 
\begin{eqnarray}
 &  & \left\langle \grave{X}^{\dagger}\vec{y},\vec{b}\right\rangle \nonumber \\
 & = & \sum_{k}e^{-j2\pi f(k-1)}\sum_{t}s_{t,k}^{*}\sum_{r}y_{r,k}b_{r,t,k}^{*}.\label{eq:tv_eq:tv_xyb-f-sum}
\end{eqnarray}
The second term can be calculated as
\begin{eqnarray}
 &  & \left(\grave{X}^{\dagger}\vec{y}\right)^{\dagger}A\left(\grave{X}^{\dagger}\vec{y}\right)\nonumber \\
 & = & \sum_{t_{1},r_{1},k_{1},t_{2},r_{2},k_{2}}s_{t_{1},k_{1}}e^{j2\pi f_{r_{1}}(k_{1}-1)}y_{r_{1},k_{1}}^{*}\nonumber \\
 &  & a_{r_{1},t_{1},k_{1},r_{2},t_{2},k_{2}}s_{t_{2},k_{2}}^{*}e^{-j2\pi f_{r_{2}}(k_{2}-1)}y_{r_{2},k_{2}}\nonumber \\
 & = & \sum_{r_{1},r_{2}}\sum_{k_{1},k_{2}}e^{j2\pi\left(f_{r_{1}}(k_{1}-1)-f_{r_{2}}(k_{2}-1)\right)}\nonumber \\
 &  & y_{r_{1},k_{1}}^{*}y_{r_{2},k_{2}}\sum_{t_{1},t_{2}}s_{t_{1},k_{1}}s_{t_{2},k_{2}}^{*}a_{r_{1},t_{1},k_{1},r_{2},t_{2},k_{2}}\nonumber \\
 & = & \sum_{r_{1}}\sum_{k_{1}}\left|y_{r_{1},k_{1}}^{*}\right|^{2}\sum_{t_{1},t_{2}}s_{t_{1},k_{1}}s_{t_{2},k_{1}}^{*}a_{r_{1},t_{1},k_{1},r_{1},t_{2},k_{1}}+\nonumber \\
 &  & 2\Re[\sum_{k_{1}=k_{2}}\sum_{r_{1}<r_{2}}e^{j2\pi(f_{r_{1}}-f_{r_{2}})(k_{1}-1)}\nonumber \\
 &  & y_{r_{1},k_{1}}^{*}y_{r_{2},k_{1}}\sum_{t_{1},t_{2}}s_{t_{1},k_{1}}s_{t_{2},k_{1}}^{*}a_{r_{1},t_{1},k_{1},r_{2},t_{2},k_{1}}+\nonumber \\
 &  & \sum_{k_{1}<k_{2}}\sum_{r_{1},r_{2}}e^{j2\pi\left(f_{r_{1}}(k_{1}-1)-f_{r_{2}}(k_{2}-1)\right)}\nonumber \\
 &  & y_{r_{1},k_{1}}^{*}y_{r_{2},k_{2}}\sum_{t_{1},t_{2}}s_{t_{1},k_{1}}s_{t_{2},k_{2}}^{*}a_{r_{1},t_{1},k_{1},r_{2},t_{2},k_{2}}].\label{eq:tv_xya-fr-sum}
\end{eqnarray}
If $f_{r}=f$, 
\begin{eqnarray*}
 &  & \left(\grave{X}^{\dagger}\vec{y}\right)^{\dagger}A\left(\grave{X}^{\dagger}\vec{y}\right)\\
 & = & \sum_{k_{1},k_{2}}e^{j2\pi f(k_{1}-k_{2})}\sum_{r_{1},r_{2}}\\
 &  & y_{r_{1},k_{1}}^{*}y_{r_{2},k_{2}}\sum_{t_{1},t_{2}}s_{t_{1},k_{1}}s_{t_{2},k_{2}}^{*}a_{r_{1},t_{1},k_{1},r_{2},t_{2},k_{2}}\\
 & = & \sum_{k_{1}}\sum_{r_{1},r_{2}}y_{r_{1},k_{1}}^{*}y_{r_{2},k_{1}}\\
 &  & \sum_{t_{1},t_{2}}s_{t_{1},k_{1}}s_{t_{2},k_{1}}^{*}a_{r_{1},t_{1},k_{1},r_{2},t_{2},k_{1}}+\\
 &  & 2\Re[\sum_{k_{1},k_{2}:k_{1}>k_{2}}e^{j2\pi f(\underset{k}{\underbrace{k_{1}-k_{2}}})}\sum_{r_{1},r_{2}}\\
 &  & y_{r_{1},k_{1}}^{*}y_{r_{2},k_{2}}\sum_{t_{1},t_{2}}s_{t_{1},k_{1}}s_{t_{2},k_{2}}^{*}a_{r_{1},t_{1},k_{1},r_{2},t_{2},k_{2}}]
\end{eqnarray*}
\begin{eqnarray}
 & = & \sum_{k_{1}}\sum_{r_{1},r_{2}}y_{r_{1},k_{1}}^{*}y_{r_{2},k_{1}}\nonumber \\
 &  & \sum_{t_{1},t_{2}}s_{t_{1},k_{1}}s_{t_{2},k_{1}}^{*}a_{r_{1},t_{1},k_{1},r_{2},t_{2},k_{1}}+\nonumber \\
 &  & 2\Re[\sum_{k=1}^{n-1}e^{j2\pi fk}\sum_{k_{1}=k+1}^{n}\sum_{r_{1},r_{2}}y_{r_{1},k_{1}}^{*}y_{r_{2},k_{1}-k}\nonumber \\
 &  & \sum_{t_{1},t_{2}}s_{t_{1},k_{1}}s_{t_{2},k_{1}-k}^{*}a_{r_{1},t_{1},k_{1},r_{2},t_{2},k_{1}-k}].\label{eq:tv_xya-f-sum}
\end{eqnarray}

\begin{comment}
\begin{eqnarray}
(\mathbf{A}+\mathbf{U}\mathbf{C}\mathbf{V})^{-1} & = & \mathbf{A}^{-1}-\mathbf{A}^{-1}\mathbf{U}(\mathbf{C}^{-1}+\mathbf{V}\mathbf{A}^{-1}\mathbf{U})^{-1}\mathbf{V}\mathbf{A}^{-1},\label{eq:tv_Woodbury-identity}
\end{eqnarray}
\begin{eqnarray}
\mathbf{BC(ABC+D)}^{-1} & = & (\mathbf{C}\mathbf{D}^{-1}\mathbf{A}+\mathbf{B}^{-1})^{-1}\mathbf{C}\mathbf{D}^{-1},\label{eq:tv_matrix_equality1}
\end{eqnarray}
\end{comment}

\section{\label{sec:tv_Cal-dgdf}Proof of Theorem \ref{thm:tv_dgdf-0} }

We find the derivatives of the three terms in 
\begin{eqnarray*}
\frac{dg(\vec{y},\vec{f})}{d\vec{f}} & = & \left[\frac{\partial g(\vec{y},\vec{f})}{\partial f_{r}}\right]_{r}
\end{eqnarray*}
 of \prettyref{eq:tv_gyf} as follows. We also consider the special
case of $f_{r}=f$.%
\begin{comment}
Calculate 
\begin{eqnarray*}
 &  & \frac{\partial\grave{X}^{\dagger}}{\partial f_{r}}\\
 & = & \left[\left[\left[\frac{\partial e^{-j2\pi\boldsymbol{f}_{r'}(k-1)}}{\partial f_{r}}s_{t,k}^{*}\right]_{1,t}\right]_{k,k}\right]_{r',r'}\\
 & = & -j2\pi\left[\frac{\partial f_{r'}}{\partial f_{r}}\left[(k-1)\left[e^{-j2\pi\boldsymbol{f}_{r'}(k-1)}s_{t,k}^{*}\right]_{1,t}\right]_{k,k}\right]_{r',r'}\\
 & = & -j2\pi\left[\frac{\partial f_{r'}}{\partial f_{r}}\left[(k-1)\right]_{k,k}\right]_{r',r'}\grave{X}^{\dagger}
\end{eqnarray*}
 first.
\end{comment}
{} The first one is %
\begin{comment}
$\underset{\grave{\boldsymbol{X}}}{\underbrace{\left[\left[\left[e^{j2\pi\boldsymbol{f}_{r,t}(k-1)}s_{t,k}\right]_{1,t}\right]_{k,k}\right]_{r,r}}}$
\end{comment}
\begin{eqnarray}
 &  & \frac{\partial2\Re\left[\left\langle \grave{X}^{\dagger}\vec{y},\vec{b}\right\rangle \right]}{\partial f_{r}}\nonumber \\
 & = & 2\Re\left[\left\langle \frac{\partial}{\partial f_{r}}\left[\left[\left[s_{t,k}^{*}\right]_{t}e^{-j2\pi f_{r'}(k-1)}y_{r',k}\right]_{k}\right]_{r'},\vec{b}\right\rangle \right]\nonumber \\
 & = & 2\Re\left[-j2\pi\left\langle \left[\left[s_{t,k}^{*}\right]_{t}(k-1)e^{-j2\pi f_{r}(k-1)}y_{r,k}\right]_{k},\left[\left[b_{r,t,k}\right]_{t}\right]_{k}\right\rangle \right]\nonumber \\
 & = & 4\pi\Im\left[\left\langle \left[\left[s_{t,k}^{*}\right]_{t}(k-1)e^{-j2\pi f_{r}(k-1)}y_{r,k}\right]_{k},\left[\left[b_{r,t,k}\right]_{t}\right]_{k}\right\rangle \right]\nonumber \\
 & = & 4\pi\Im\left[\sum_{k}e^{-j2\pi f_{r}(k-1)}(k-1)\sum_{t}s_{t,k}^{*}y_{r,k}b_{r,t,k}^{*}\right]\nonumber \\
 & = & -4\pi\Im\left[\sum_{k=1}^{n-1}e^{j2\pi f_{r}k}k\sum_{t}s_{t,k+1}y_{r,k+1}^{*}b_{r,t,k+1}\right].\label{eq:tv_d1-df-1}
\end{eqnarray}

If $f_{r}=f$, then 
\begin{eqnarray}
 &  & \frac{\partial2\Re\left[\left\langle \grave{X}^{\dagger}\vec{y},\vec{b}\right\rangle \right]}{\partial f}\nonumber \\
 & = & \sum_{r}\left.\frac{\partial2\Re\left[\left\langle \grave{X}^{\dagger}\vec{y},\vec{b}\right\rangle \right]}{\partial f_{r}}\right|_{f_{r}=f}\nonumber \\
 & = & -4\pi\Im\left[\sum_{k=1}^{n-1}e^{j2\pi fk}k\sum_{r,t}s_{t,k+1}y_{r,k+1}^{*}b_{r,t,k+1}\right].\label{eq:tv_d1-df-f}
\end{eqnarray}

From (\ref{eq:tv_xya-fr-sum}), the second term is calculated as
\begin{eqnarray*}
 &  & \frac{\partial\left(\grave{X}^{\dagger}\vec{y}\right)^{\dagger}A\left(\grave{X}^{\dagger}\vec{y}\right)}{\partial f_{r}}\\
 & = & -4\pi\Im[\sum_{k_{1}}(k_{1}-1)\sum_{r_{1,}r_{2}:r_{1}<r_{2}}\frac{\partial(f_{r_{1}}-f_{r_{2}})}{\partial f_{r}}\\
 &  & e^{j2\pi(f_{r_{1}}-f_{r_{2}})(k_{1}-1)}\eta_{r_{1},k_{1},r_{2},k_{1}}+\\
 &  & \sum_{k_{1},k_{2}:k_{1}<k_{2}}\sum_{r_{1},r_{2}}\frac{\partial\left(f_{r_{1}}(k_{1}-1)-f_{r_{2}}(k_{2}-1)\right)}{\partial f_{r}}\\
 &  & e^{j2\pi\left(f_{r_{1}}(k_{1}-1)-f_{r_{2}}(k_{2}-1)\right)}\eta_{r_{1},k_{1},r_{2},k_{2}}]
\end{eqnarray*}
\begin{eqnarray*}
 & = & -4\pi\Im[\sum_{k_{1}}(k_{1}-1)(\\
 &  & \sum_{r_{2}:r<r_{2}}e^{j2\pi(f_{r}-f_{r_{2}})(k_{1}-1)}\eta_{r,k_{1},r_{2},k_{1}}-\\
 &  & \sum_{r_{1}:r_{1}<r}e^{j2\pi(f_{r_{1}}-f_{r})(k_{1}-1)}\eta_{r_{1},k_{1},r,k_{1}})+\\
 &  & \sum_{k_{1},k_{2}:k_{1}<k_{2}}(\sum_{r_{2}}(k_{1}-1)\\
 &  & e^{j2\pi\left(f_{r}(k_{1}-1)-f_{r_{2}}(k_{2}-1)\right)}\eta_{r,k_{1},r_{2},k_{2}}-\\
 &  & \sum_{r_{1}}(k_{2}-1)\\
 &  & e^{j2\pi\left(f_{r_{1}}(k_{1}-1)-f_{r}(k_{2}-1)\right)}\eta_{r_{1},k_{1},r,k_{2}})]
\end{eqnarray*}
\begin{eqnarray}
 & = & -4\pi\Im[\sum_{k_{1}}(k_{1}-1)(\nonumber \\
 &  & \sum_{r_{2}:r<r_{2}}e^{j2\pi(f_{r}-f_{r_{2}})(k_{1}-1)}\eta_{r,k_{1},r_{2},k_{1}}-\nonumber \\
 &  & \sum_{r_{1}:r_{1}<r}e^{j2\pi(f_{r_{1}}-f_{r})(k_{1}-1)}\eta_{r_{1},k_{1},r,k_{1}})+\nonumber \\
 &  & \sum_{k_{1},k_{2}:k_{1}<k_{2}}(\sum_{r_{2}}(k_{1}-1)\nonumber \\
 &  & e^{j2\pi\left(f_{r}(k_{1}-1)-f_{r_{2}}(k_{2}-1)\right)}\eta_{r,k_{1},r_{2},k_{2}}-\nonumber \\
 &  & \sum_{r_{1}}(k_{2}-1)\nonumber \\
 &  & e^{j2\pi\left(f_{r_{1}}(k_{1}-1)-f_{r}(k_{2}-1)\right)}\eta_{r_{1},k_{1},r,k_{2}})],\label{eq:tv_d2-dfr}
\end{eqnarray}
where
\begin{eqnarray*}
\eta_{r_{1},k_{1},r_{2},k_{2}} & \triangleq & y_{r_{1},k_{1}}^{*}y_{r_{2},k_{2}}\sum_{t_{1},t_{2}}s_{t_{1},k_{1}}s_{t_{2},k_{2}}^{*}a_{r_{1},t_{1},k_{1},r_{2},t_{2},k_{2}}\\
 & = & \eta_{r_{2},k_{2},r_{1},k_{1}}^{*}.
\end{eqnarray*}
{} If $f_{r}=f$, then from (\ref{eq:tv_xya-f-sum}),
\begin{eqnarray}
 &  & \frac{\partial\left(\grave{X}^{\dagger}\vec{y}\right)^{\dagger}A\left(\grave{X}^{\dagger}\vec{y}\right)}{\partial f}\nonumber \\
 & = & 2\Re[\sum_{k=1}^{n-1}\frac{\partial}{\partial f}e^{j2\pi fk}\sum_{k_{1}=k+1}^{n}\sum_{r_{1},r_{2}}y_{r_{1},k_{1}}^{*}y_{r_{2},k_{1}-k}\nonumber \\
 &  & \sum_{t_{1},t_{2}}s_{t_{1},k_{1}}s_{t_{2},k_{1}-k}^{*}a_{r_{1},t_{1},k_{1},r_{2},t_{2},k_{1}-k}].\nonumber \\
 & = & -4\pi\Im[\sum_{k=1}^{n-1}e^{j2\pi fk}k\sum_{k_{1}=k+1}^{n}\sum_{r_{1},r_{2}}y_{r_{1},k_{1}}^{*}y_{r_{2},k_{1}-k}\nonumber \\
 &  & \sum_{t_{1},t_{2}}s_{t_{1},k_{1}}s_{t_{2},k_{1}-k}^{*}a_{r_{1},t_{1},k_{1},r_{2},t_{2},k_{1}-k}].\label{eq:tv_d2-df}
\end{eqnarray}

\begin{comment}
{[}This method used partial derivative and is harder to simplify.
If $f_{r}=f$, then
\begin{eqnarray}
 &  & \frac{\partial\left(\grave{X}^{\dagger}\vec{y}\right)^{\dagger}A\left(\grave{X}^{\dagger}\vec{y}\right)}{\partial f}\nonumber \\
 & = & -4\pi\Im[\sum_{k_{1},k_{2}}(k_{1}-1)e^{j2\pi f\left(k_{1}-k_{2}\right)}\sum_{r_{1},r_{2}}\nonumber \\
 &  & y_{r,k_{1}}^{*}y_{r_{2},k_{2}}\sum_{t_{1},t_{2}}s_{t_{1},k_{1}}s_{t_{2},k_{2}}^{*}a_{r,t_{1},k_{1},r_{2},t_{2},k_{2}}].\label{eq:tv_d2-df-f}
\end{eqnarray}
\end{comment}

The third term is 
\begin{eqnarray}
 &  & \frac{\partial-\frac{1}{2}\left(\vec{f}-\vec{\mu}_{\vec{\boldsymbol{f}}}\right)^{\dagger}\Sigma_{\vec{\boldsymbol{f}}}^{-1}\left(\vec{f}-\vec{\mu}_{\vec{\boldsymbol{f}}}\right)}{\partial f_{r}}\nonumber \\
 & = & -\left(\frac{\partial}{\partial f_{r}}\left(\vec{f}-\vec{\mu}_{\vec{\boldsymbol{f}}}\right)\right)^{\dagger}\Sigma_{\vec{\boldsymbol{f}}}^{-1}\left(\vec{f}-\vec{\mu}_{\vec{\boldsymbol{f}}}\right)\nonumber \\
 & = & -\sum_{r_{1},r_{2}}\frac{\partial f_{r_{1}}}{\partial f_{r}}\varsigma_{r_{1},r_{2}}(f_{r_{2}}-\mu_{\boldsymbol{f}_{r_{2}}})\nonumber \\
 & = & -\sum_{r_{2}}\varsigma_{r,r_{2}}(f_{r_{2}}-\mu_{\boldsymbol{f}_{r_{2}}}),\label{eq:tv_d3-df-1}
\end{eqnarray}
where $\Sigma_{\vec{\boldsymbol{f}}}^{-1}=\left[\varsigma_{r_{1},r_{2}}\right]_{r_{1},r_{2}}$.
If the $\boldsymbol{f}_{r}$'s are independent, 
\begin{eqnarray}
 &  & \frac{\partial-\frac{1}{2}\left(\vec{f}-\vec{\mu}_{\vec{\boldsymbol{f}}}\right)^{\dagger}\Sigma_{\vec{\boldsymbol{f}}}^{-1}\left(\vec{f}-\vec{\mu}_{\vec{\boldsymbol{f}}}\right)}{\partial f_{r}}\nonumber \\
 & = & -\sigma_{\boldsymbol{f}_{r}}^{2}(f_{r}-\mu_{\boldsymbol{f}_{r}}).\label{eq:tv_d3-dfr-independent}
\end{eqnarray}
If $f_{r}=f$, then
\begin{eqnarray}
 &  & -\frac{1}{2}\sigma_{\boldsymbol{f}}^{-2}\frac{\partial}{\partial f}\left|f-\mu_{\boldsymbol{f}_{\delta}}\right|^{2}\nonumber \\
 & = & -\sigma_{\boldsymbol{f}}^{-2}\left(f-\mu_{\boldsymbol{f}}\right).\label{eq:tv_d3-df-f}
\end{eqnarray}

Combing eq. (\ref{eq:tv_d1-df-1}, \ref{eq:tv_d2-dfr}, \ref{eq:tv_d3-dfr-independent}),
we obtain eq. (\ref{eq:tv_dg-df-fr}). Combing eq. (\ref{eq:tv_d1-df-f},
\ref{eq:tv_d2-df}, \ref{eq:tv_d3-df-f}), we obtain eq. (\ref{eq:tv_dg-df-f}).
, \ref{eq:tv_rtheta-general}).

\section{\label{sec:tv_Proof-CRLB}Proof of Theorem \ref{thm:tv_CRLB} of
the Cramer-Rao Lower Bound \label{sec:tv_Proof-of-Theorem}}

For the case of $\boldsymbol{f}_{r}=\boldsymbol{f}$, we calculate
the BCRLB. First calculate
\begin{eqnarray}
 &  & \frac{\partial^{2}\ln\left(f_{\vec{\boldsymbol{y}}|\boldsymbol{f}}(\vec{y}|f)f_{\boldsymbol{f}}(f)\right)}{\partial f^{2}}\nonumber \\
 & = & \frac{\partial^{2}g(\vec{y},f)}{\partial f^{2}}\nonumber \\
 & = & -\frac{\partial}{\partial f}4\pi\Im\left[\sum_{k=1}^{n-1}e^{j2\pi fk}kz_{k}\right]\nonumber \\
 &  & -\frac{\partial}{\partial f}\sigma_{\boldsymbol{f}}^{-2}\left(f-\mu_{\boldsymbol{f}}\right)\label{eq:tv_used-df-dfd}\\
 & = & -4\pi\Im\left[j2\pi\sum_{k=1}^{n-1}e^{j2\pi fk}k^{2}z_{k}\right]-\sigma_{\boldsymbol{f}}^{-2}\nonumber \\
 & = & -8\pi^{2}\Re\left[\sum_{k=1}^{n-1}e^{j2\pi fk}k^{2}z_{k}\right]-\sigma_{\boldsymbol{f}}^{-2},\nonumber 
\end{eqnarray}
\begin{comment}
\begin{eqnarray}
0 & = & \frac{\partial g(\vec{y},f)}{\partial f}\nonumber \\
 & = & -4\pi\Im\left[\sum_{k=1}^{n-1}e^{j2\pi fk}kz_{k}\right]\nonumber \\
 &  & -\sigma_{\boldsymbol{f}}^{-2}\left(f-\vec{\mu}_{\boldsymbol{f}}\right),\label{eq:tv_dg-df-f-1}
\end{eqnarray}
\end{comment}
where we have used \prettyref{eq:tv_dg-df-f}. 

Note that $\text{E}\left[\cdot\right]=\text{E}_{\boldsymbol{f}}\left[\text{E}_{\vec{\boldsymbol{y}}|\boldsymbol{f}}\left[\cdot\right]\right]$.
We calculate 
\begin{eqnarray}
 &  & \text{E}_{\vec{\boldsymbol{y}}|\boldsymbol{f}}\left[\frac{\partial^{2}\ln\left(f_{\vec{\boldsymbol{y}}|\boldsymbol{f}}(\vec{\boldsymbol{y}}|\boldsymbol{f})f_{\boldsymbol{f}}(\boldsymbol{f})\right)}{\partial\boldsymbol{f^{2}}}\right]\nonumber \\
 & = & -8\pi^{2}\Re\left[\sum_{k=1}^{n-1}e^{j2\pi\boldsymbol{f}k}k^{2}\text{E}_{\vec{\boldsymbol{y}}|\boldsymbol{f}}\left[\boldsymbol{z}_{k}\right]\right]-\sigma_{\boldsymbol{f}}^{-2}\label{eq:tv_Edln}
\end{eqnarray}
 first. Inspecting \prettyref{eq:tv_rtheta-general}, %
\begin{comment}
\begin{eqnarray}
z_{k} & \triangleq & r_{k}e^{-j\theta_{k}}\nonumber \\
 & \triangleq & \sum_{r,t}s_{t,k+1}y_{r,k+1}^{*}b_{r,t,k+1}+\nonumber \\
 &  & \sum_{k_{1}=k+1}^{n}\sum_{r_{1},t_{1},r_{2},t_{2}}a_{r_{1},t_{1},k_{1},r_{2},t_{2},k_{1}-k}\times\nonumber \\
 &  & s_{t_{1},k_{1}}s_{t_{2},k_{1}-k}^{*}y_{r_{2},k_{1}-k}y_{r_{1},k_{1}}^{*}.\label{eq:tv_rtheta-general-2}
\end{eqnarray}
\end{comment}
we need to calculate %
\begin{comment}
\begin{eqnarray*}
\boldsymbol{y}_{r,k} & = & \sum_{t=1}^{l_{\text{t}}}e^{j2\pi\boldsymbol{f}_{r,t}(k-1)}s_{t,k}\boldsymbol{h}_{r,t,k}+\boldsymbol{n}_{r,k},
\end{eqnarray*}
\end{comment}
\begin{eqnarray*}
 &  & \text{E}_{\vec{\boldsymbol{y}}|\boldsymbol{f}}\left[\boldsymbol{y}_{r,k+1}^{*}\right]\\
 & = & \text{E}_{\vec{\boldsymbol{y}}|\boldsymbol{f}}\left[e^{-j2\pi\boldsymbol{f}k}\sum_{t'}s_{t',k+1}^{*}\boldsymbol{h}_{r,t',k+1}^{*}+\boldsymbol{n}_{r,k+1}^{*}\right]\\
 & = & e^{-j2\pi\boldsymbol{f}k}\sum_{t'}s_{t',k+1}^{*}\mu_{\boldsymbol{h}_{r,t',k+1}}^{*}
\end{eqnarray*}
{} and 
\begin{eqnarray*}
 &  & \text{E}_{\vec{\boldsymbol{y}}|\boldsymbol{f}}\left[y_{r_{1},k_{1}}^{*}y_{r_{2},k_{2}}\right]\\
 & = & \text{E}_{\vec{\boldsymbol{y}}|\boldsymbol{f}}\left[\left(e^{-j2\pi\boldsymbol{f}(k_{1}-1)}\sum_{t_{1}'}s_{t_{1}',k_{1}}^{*}\boldsymbol{h}_{r_{1},t_{1}',k_{1}}^{*}+\boldsymbol{n}_{r_{1},k_{1}}^{*}\right)\right.\\
 &  & \left.\left(e^{j2\pi\boldsymbol{f}(k_{2}-1)}\sum_{t_{2}'}s_{t_{2}',k_{2}}\boldsymbol{h}_{r_{2},t_{2}',k_{2}}+\boldsymbol{n}_{r_{2},k_{2}}\right)\right]\\
 & \stackrel{k_{2}=k_{1}-k}{=} & e^{-j2\pi\boldsymbol{f}k}\sum_{t_{1}'}s_{t_{1}',k_{1}}^{*}\sum_{t_{2}'}s_{t_{2}',k_{1}-k}\text{E}\left[\boldsymbol{h}_{r_{1},t_{1}',k_{1}}^{*}\boldsymbol{h}_{r_{2},t_{2}',k_{1}-k}\right]\\
 &  & +\delta[r_{1}-r_{2}]\delta[k],
\end{eqnarray*}
 where $\text{E}\left[\boldsymbol{h}_{r_{1},t_{1}',k_{1}}^{*}\boldsymbol{h}_{r_{2},t_{2}',k_{2}}\right]=c_{\boldsymbol{h}_{r_{1},t_{1}',k_{1}},\boldsymbol{h}_{r_{2},t_{2}',k_{2}}}^{*}+\mu_{\boldsymbol{h}_{r_{2},t_{2}',k_{2}}}\mu_{\boldsymbol{h}_{r_{1},t_{1}',k_{1}}}^{*}$;
and $c_{\boldsymbol{h}_{r_{1},t_{1}',k_{1}},\boldsymbol{h}_{r_{2},t_{2}',k_{2}}}^{*}$
is defined in \prettyref{eq:tv_cov-h}. They are used to obtain%
\begin{comment}
\begin{eqnarray}
r_{k}e^{-j\theta_{k}} & \triangleq & \sum_{r,t}s_{t,k+1}y_{r,k+1}^{*}b_{r,t}+\nonumber \\
 &  & \sum_{k_{1}=k+1}^{n}\sum_{r_{1},t_{1},r_{2},t_{2}}a_{r_{1},t_{1},r_{2},t_{2}}\times\nonumber \\
 &  & s_{t_{1},k_{1}}s_{t_{2},k_{1}-k}^{*}y_{r_{2},k_{1}-k}y_{r_{1},k_{1}}^{*}.\label{eq:tv_rtheta-general-1}
\end{eqnarray}
\end{comment}
{} 
\begin{eqnarray}
 &  & \text{E}_{\vec{\boldsymbol{y}}|\boldsymbol{f}}\left[\boldsymbol{z}_{k}\right]\nonumber \\
 & = & e^{-j2\pi\boldsymbol{f}k}\sum_{r,t}\sum_{t'}s_{t,k+1}s_{t',k+1}^{*}\mu_{\boldsymbol{h}_{r,t',k+1}}^{*}b_{r,t,k+1}+\nonumber \\
 &  & e^{-j2\pi\boldsymbol{f}k}\sum_{k_{1}=k+1}^{n}\sum_{r_{1},t_{1},r_{2},t_{2}}a_{r_{1},t_{1},k_{1},r_{2},t_{2},k_{1}-k}\times\nonumber \\
 &  & s_{t_{1},k_{1}}s_{t_{2},k_{1}-k}^{*}\sum_{t_{2}'}s_{t_{2}',k_{1}-k}\sum_{t_{1}'}s_{t_{1}',k_{1}}^{*}\times\nonumber \\
 &  & \left(c_{\boldsymbol{h}_{r_{1},t_{1}',k_{1}},\boldsymbol{h}_{r_{2},t_{2}',k_{1}-k}}^{*}+\mu_{\boldsymbol{h}_{r_{2},t_{2}',k_{1}-k}}\mu_{\boldsymbol{h}_{r_{1},t_{1}',k_{1}}}^{*}\right),\ k\ne0.\label{eq:tv_Ertheta}
\end{eqnarray}
Plug \prettyref{eq:tv_Ertheta} into \prettyref{eq:tv_Edln}, we see
that $\text{E}_{\vec{\boldsymbol{y}}|\boldsymbol{f}}\left[\frac{\partial^{2}\ln\left(f_{\vec{\boldsymbol{y}}|\boldsymbol{f}}(\vec{\boldsymbol{y}}|\boldsymbol{f})f_{\boldsymbol{f}}(\boldsymbol{f})\right)}{\partial\boldsymbol{f^{2}}}\right]$
is not a function of $\boldsymbol{f}$. Therefore, 
\begin{eqnarray*}
 &  & \text{E}_{\vec{\boldsymbol{y}}|\boldsymbol{f}}\left[\frac{\partial^{2}\ln\left(f_{\vec{\boldsymbol{y}}|\boldsymbol{f}}(\vec{\boldsymbol{y}}|\boldsymbol{f})f_{\boldsymbol{f}}(\boldsymbol{f})\right)}{\partial\boldsymbol{f}^{2}}\right]\\
 & = & \text{E}_{\boldsymbol{f}}\left[\text{E}_{\vec{\boldsymbol{y}}|\boldsymbol{f}}\left[\frac{\partial^{2}\ln\left(f_{\vec{\boldsymbol{y}}|\boldsymbol{f}}(\vec{\boldsymbol{y}}|\boldsymbol{f})f_{\boldsymbol{f}}(\boldsymbol{f})\right)}{\partial\boldsymbol{f}^{2}}\right]\right],
\end{eqnarray*}
which is plugged into \prettyref{eq:tv_1-E} to obtain $\beta$ of
BCRLB in \prettyref{eq:tv_-E-general}.

One can calculate $\text{E}_{\vec{\boldsymbol{y}}|\{\boldsymbol{f}=f\}}\left[\frac{\partial^{2}\ln\left(f_{\vec{\boldsymbol{y}}|\boldsymbol{f}}(\vec{\boldsymbol{y}}|f)\right)}{\partial f^{2}}\right]$
and observe that it is obtained by setting $\sigma_{\boldsymbol{f}}^{-2}=0$.
This provides the CRLB.